\documentclass[10pt,letterpaper]{article}
\usepackage[top=0.85in,left=2.75in,footskip=0.75in]{geometry}

\usepackage{amsmath,amssymb}

\usepackage{changepage}

\usepackage[utf8x]{inputenc}

\usepackage{textcomp,marvosym}

\usepackage{cite}

\usepackage{nameref,hyperref}

\usepackage[right]{lineno}

\usepackage{microtype}
\DisableLigatures[f]{encoding = *, family = * }

\usepackage[table]{xcolor}

\usepackage{array}

\usepackage{rotating}
\usepackage{changes}
\usepackage{longtable}
\usepackage{outlines}
\usepackage{titlesec}
\usepackage{algorithm}
\usepackage{algpseudocode}
\usepackage{cleveref}
\usepackage{subfig}
\usepackage{url}

\newcolumntype{+}{!{\vrule width 2pt}}

\newlength\savedwidth

\raggedright
\usepackage[aboveskip=1pt,labelfont=bf,labelsep=period,justification=raggedright,singlelinecheck=off]{caption}

\makeatletter
\renewcommand{\@biblabel}[1]{\quad#1.}
\makeatother

\date{}

\usepackage{lastpage,fancyhdr,graphicx}
\usepackage{epstopdf}
\fancyheadoffset[L]{2.25in}
\fancyfootoffset[L]{2.25in}

\newcommand{\bbP}{\mathbb{P}}
\newcommand{\xin}{\xi_{\rm noise}}
\newcommand{\xis}{\xi_{\rm sys}}
\DeclareMathOperator*{\argmax}{argmax}
\DeclareMathOperator*{\argmin}{argmin}

\begin{document}
\vspace*{0.2in}

\begin{flushleft}
{\Large
\textbf\newline{Offline and online data assimilation for real-time blood glucose forecasting in type 2 diabetes} 
}
\newline
\\
Matthew E Levine\textsuperscript{1*},
George Hripcsak\textsuperscript{1},
Lena Mamykina\textsuperscript{1},
Andrew Stuart\textsuperscript{2\ddag},
David J Albers\textsuperscript{1\ddag}
\\
\bigskip
\textbf{1} Department of Biomedical Informatics, Columbia University, New York, NY, USA
\\
\textbf{2} Department of Computing and Mathematical Sciences, California Institute of Technology, Pasadena, CA, USA
\\
\bigskip

%
%

\ddag These authors contributed equally to this work.




* mel2193@cumc.columbia.edu

\end{flushleft}

\section{Abstract}
We evaluate the benefits of combining different offline and online data assimilation methodologies to improve personalized blood glucose prediction with type 2 diabetes self-monitoring data.
We collect self-monitoring data (nutritional reports and pre- and post-prandial glucose measurements) from 4 individuals with diabetes and 2 individuals without diabetes.
We write \emph{online} to refer to methods that update state and parameters sequentially as nutrition and glucose data are received, and \emph{offline} to refer to methods that estimate parameters over a fixed data set, distributed over a time window containing multiple nutrition and glucose measurements.

We fit a model of ultradian glucose dynamics to the first half of each data set using offline (MCMC and nonlinear least squared optimization) and online (unscented Kalman filter and an unfiltered model---a dynamical model driven by nutrition data that does not update states) data assimilation methods. 
Model parameters estimated over the first half of the data are used within online forecasting methods to issue forecasts over the second half of each data set. Prediction performance is measured with common model selection criteria, as well as a diabetes-specific metric that weights errors by clinical importance.

Offline data assimilation methods provided consistent advantages in predictive performance and practical usability in 4 of 6 patient data sets compared to online data assimilation methods alone; yet 2 of 6 patients were best predicted with a strictly online approach.
Interestingly, parameter estimates generated offline led to worse predictions when fed to a stochastic filter than when used in a simple, unfiltered model that incorporates new nutritional information, but does not update model states based on glucose measurements.

The relative improvements seen from the unfiltered model, when carefully trained offline, exposes challenges in model sensitivity and filtering applications, but also opens the possibility for improved glucose forecasting and relaxing patients' self-monitoring requirements.


\section{Author summary}
Type 2 diabetes is currently one of the most prevalent and costly chronic health conditions in the world, but the growing availability of self-monitoring data gives hope for personalized interventions that can help people with diabetes and their clinicians improve treatments and self-management strategies. Here, we investigate novel combinations of methodologies for both inferring and predicting glucose dynamics in the context of a simple glucose-insulin model and real-world type 2 diabetes data. We find that offline training on 40 blood glucose measurements and roughly 20 meal records was sufficient to outperform previously established methods for this same data and model.

\section{Introduction}
Type 2 diabetes is currently one of the most prevalent and costly chronic health conditions; it affects over $8\%$ of American adults and $95\%$ of all individuals with diabetes~\cite{ADA_standards_2014}. Recent research suggests that successful diabetes treatment and self-management ought to be tailored not only to individuals' behaviors, attitudes, and goals, but also personalized to their unique physiology. However, it is difficult for individuals and clinicians alike to predict the short-term clinical impact of daily self-management and nutritional choices~\cite{mamykina_data-driven_2016}. Nevertheless, the growing availability of self-monitoring data gives hope for computational inference of important relationships in diabetes self-management that can support decision-making. Here we study how computational techniques in data assimilation, inverse problems, and physiologic modeling can be leveraged to produce high-fidelity forecasts of post-meal blood glucose levels, given only sparse, noisy, and biased data that are collected in free-living conditions. 



In order to frame progress toward improved glycemic forecasting, we conceptualize the levers that the glucose forecast is sensitive to, then work toward a better understanding along one or more of these directions. The key levers that impact effective glycemic forecasting include: i) data, or the type and quality of data that are available (e.g. finger sticks or continuous monitors, nutrition, activity); ii) modeling technology, or the models used to power the blood glucose forecast; iii) inference methodology, or the choice of inference scheme (e.g. data assimilation), used to leverage the model(s) and generate a forecast, iv) evaluation and uncertainty quantification, or the criteria used to evaluate the forecasts, parameter estimates, and model choices, and v) translation of the forecast, or the constructs used to communicate forecasts effectively (e.g. graphical presentations, qualitative explanations, long-term risk scores).
 
Here we focus on the sensitivity of blood glucose forecasting to methods of inference (i.e. lever (iii)) over parameters and states of a particular model. We use three different evaluation metrics to characterize how choices in inference methodology can affect blood glucose forecasting. We use a single physiologic model to simplify the analysis, we use data constrained by the reality of self-monitoring in type 2 diabetes, and we do not focus on approaches to using the forecasts to improve self-management of diabetes.  We consider online and offline data assimilation situations, where 
\emph{online} refers to methods that update states and parameters sequentially as each nutrition and glucose data point is received, and \emph{offline} refers to methods that estimate parameters over a fixed data set, distributed over an entire window of time in which multiple nutrition and glucose data measurements are received. Methods that are extremely computationally intensive, for example, may only be able to be used offline.


We use online methods that allow glucose forecasts to be sequentially generated and adapted in-the-moment (e.g. during meal-time), typically by using a stochastic filter. We also use offline methods to more thoroughly estimate model parameters through optimization approaches and Bayesian inverse methods (i.e. MCMC) that require significant runtime. 
Our results compare each of these approaches individually and evaluate the challenges and opportunities of fusing the two paradigms.
 
Among the many results we present, there are two primary findings: 1) We find that Markov Chain Monte Carlo parameter estimation paired with a simple unfiltered forecasting method, which is driven by nutrition, but does not update model states based on new measurements, often outperformed other methods, including optimization approaches to parameter estimation and sequential data assimilation with unscented Kalman filters. This was a surprise, given our expectation that well-estimated parameters would allow state filtering and tracking to improve, rather than degrade. 2) A strictly online data assimilation approach still remains advantageous for a significant subset of participants. Unsurprisingly, we do not observe a universally best inference scheme across all people.

\subsection{Self-management in Type 2 Diabetes}
Recent work by Zeevi et al. demonstrated that different individuals can have markedly unique glycemic responses to nutrition, which can depend on complex physiological and contextual factors~\cite{zeevi_personalized_2015}. While the American Diabetes Association (ADA) provides generic guidelines for nutrition and physical activity to people with diabetes, it also stresses the importance of personalized self-care. In fact, self-management has been shown to be a critical component of care that can help to reduce diabetes-related complications~\cite{AADE,khattab_factors_2010}.

The American Association of Diabetes Educators lists self-monitoring as one of seven essential self-management behaviors~\cite{AADE}.
While self-monitoring of blood glucose levels is a common practice in diabetes self-management, its frequency varies drastically between 9-12 times per day for individuals on insulin therapy to once daily or less for individuals who manage their diabetes with oral agents. In addition to glucose levels, self-monitoring in diabetes may include keeping track of meals, physical activity, sleep, medication, and other relevant daily activities. There exist a wide variety of electronic tools for keeping track of these activities, including digital diaries and wearable activity trackers.

In our previous work, we examined individuals' ability to use self-monitoring data to both identify patterns in glycemic response to nutrition (participants used a mobile application to capture the glycemic impact and nutritional content of each of their meals) and to predict glycemic response to future meals~\cite{mamykina_data-driven_2016}. 
In the study, individuals with diabetes and experienced diabetes educators made predictions about blood glucose levels with inconsistent accuracy, often differing substantially from true recorded measurements. 
These findings, along with others (e.g.~\cite{paterson_expert_2000}) indicate that individuals with diabetes often experience considerable challenges in understanding reasons behind fluctuations in BG levels and how this relates to their behavior.
In order to help bridge this gap, we seek to design informatics solutions using advanced data science methods that provide cognitive support for understanding the short-term glycemic impact of an individual's nutritional choices.

\subsection{Personalized glucose prediction in type 2 diabetes}

Glucose oscillations vary across people and context in a) periodicity from order minutes to hours and b) amplitude from order ten to one-hundred. 
In contrast, self-monitoring records of individuals with type 2 diabetes are often irregularly sampled, sparse, and can include several meals per day and from 1 to 12 blood glucose readings captured at different times of day.
Self-monitoring data are not only sparse with respect to underlying dynamics, but also tend to be noisy and biased. These sources of complexity present a wide array of important and challenging computational problems.   

Indeed, glucose prediction in diabetes has been studied deeply by researchers across many disciplines, and has resulted in a diversity of data-driven approaches for personalized glucose forecasting~\cite{marmarelis_data-driven_2014, funatsu_glucose_2011}. Efforts that focus on physiologic inference~\cite{vahidi_detection_2015,beverlin_algorithm_2011,vahidi_comprehensive_2015,dalla_man}) have taken a variety of approaches to identifying suitable models of a diabetic's glucose system (e.g. nonlinear weiner model~\cite{rollins_free-living_2010,beverlin_algorithm_2011,kotz_multiple-input_2014}, neural networks~\cite{zitar_towards_2005}, probabilistic models~\cite{murata_gh_probabilistic_2004}, and mechanistic systems of ordinary differential equations~\cite{dalla_man,vahidi_comprehensive_2015}).

Sturis et al.~\cite{sturis_abnormalities_1992} interrogated their model of ultradian glucose-insulin dynamics using time-series analysis methods on normal and type 2 diabetic subjects, and found that irregular coupling of insulin and glucose oscillations was a distinguishing feature of diabetics more often than oscillation frequency or amplitude. We employ the model proposed by Sturis et al.~\cite{sturisUltradian} to account for some of these important interactions that simpler models (e.g.~\cite{bergman_physiologic_1981}) do not describe.

Many of the desired prediction tasks have been approached by wrapping the aforementioned models in different predictive algorithms and inference schemes (e.g. stochastic filters~\cite{barazandegan_assessment_2014}, gaussian process models~\cite{valletta_gaussian_2009}, fuzzy logic~\cite{ekram_feedback_2012}, and many other machine learning methods~\cite{sudharsan_hypoglycemia_2015,funatsu_glucose_2011,plis_machine_2014,gibson_development_2013}). These studies have paved the way towards personalized interventions for people with diabetes, but most studies fail to show their applicability in settings where data collection is  restricted to realistic, free-living conditions of people with type 2 diabetes. 

Albers et al. used freely collected diabetes self-monitoring data to demonstrate that, given sparse, noisy data, data assimilation with physiologic modeling can: 1) generate glucose forecasts with accuracy similar to forecasts made by diabetes experts, 2) personalize to an individual with 1-2 weeks of data, and 3) respond to non-stationarities and changes in individuals' behavior and/or physiology. The approach used a dual unscented Kalman filter~\cite{julier_unscented_2004} to fit two mechanistic models of glucose-insulin dynamics ~\cite{dalla_man,sturisUltradian}), and has since been operationalized in a pilot study of a mobile application, GlucOracle, where we can test hypotheses regarding the clinical impact of personalized, in-the-moment, nutrition-based glucose forecasts.

While the previous implementation by Albers et al. relies on sequential state and parameter estimation (i.e. a prediction-correction scheme), we also recognize that opportunities exist to periodically perform more costly computations \emph{offline} that do not interfere with online forecasting demands. Offline approaches to parameter estimation may improve predictive performance by searching a higher-dimensional parameter space over larger windows of data than can be assessed using fast online methods.

\subsection{A combined offline--online approach}
The overarching goal of this work is to understand how to improve forecasting accuracy on real-world type 2 diabetes self-monitoring data. We attempt to achieve better results than the method outlined by Albers et al., which entailed an \emph{online} prediction-correction scheme for state and parameter estimation (dual unscented Kalman filter) using two popular physiologic models.
We specifically focus on the opportunities and challenges of introducing an additional class of methods, which we collectively refer to as \emph{offline} methods, that typically make fewer assumptions about the independence of data and predictions and often involve lengthy computations that cannot be relied on for in-the-moment decision making. We hypothesize that offline estimation methods will be most useful to track slow moving quantities, like biological parameters that govern the rates of physiologic processes (e.g. insulin sensitivity rates). We expect that online estimation, on the other hand, lends itself more naturally to tracking time-varying phyisiologic states (e.g. blood glucose), which can frequently endure unexpected fluctuations, due in large part to model insufficiencies. 

Moreover, we suspect that offline parameter estimation can be paired with online forecasting by periodically updating parameters of an online data assimilator to reflect improved parameter estimates acquired offline. Here, we evaluate multiple methods of offline parameter estimation in conjunction with multiple methods of online prediction, and use the ultradian glucose model~\cite{sturisUltradian} in order to determine how offline methods can be used to improve glucose forecasting. We perform this evaluation using standard model selection criteria, including mean squared error (MSE) and linear (Pearson) correlation, as well as a validated diabetes-specific metric that weights errors according to their clinical significance (Parkes Quality)~\cite{parkes_new_2000,pfutzner_technical_2013}. We also consider the relationship between performance gains, data quantity, and computational cost.

\section{Methods}

We wish to identify the best forecasting schemes for use with type 2 diabetes self-monitoring data. While there exist many options for designing forecasting methodologies, we focus on a few approaches to fitting and forecasting with a single ultradian model of glucose dynamics. We consider methods for estimating model parameters (Markov Chain Monte Carlo, Nelder-Mead optimization, and dual unscented Kalman filtering) and forecasting model states (a nutrition-driven model, either combined with an unscented Kalman filter or, once tuned to offline data, run in an unfiltered predictive mode). Moreover, we hypothesize that different methodological combinations (e.g. first, learn model parameters using Nelder-Mead optimization, then feed these parameter estimates to an unscented Kalman filter for subsequent online forecasting) will yield different levels of forecasting performance.

In order to evaluate combinations of methods for offline parameter estimation and online state forecasting, we: 1) collect diabetes self-monitoring data, 2) perform multiple types of offline parameter estimation for the first half of each patient data-set, 3) perform multiple types of online state forecasting over the second half of the patient data-set (given parameter estimates from step (2)), and 4) evaluate the forecasting performance in step (3).

\subsection{Data collection and pre-processing}

\begin{table}[!ht]
\begin{adjustwidth}{-2.25in}{0in} 
\centering
\caption{{\bf Data Summary}}
\label{table:participants}
\scriptsize
\begin{tabular}{|l|c|c|c|c|c|c|c|c|c|}
\hline
Participant ID & P1 & P2 & P3 & P4 & P5 & P6 \\ \hline 
Age & $40-50$ & $40-50$ & $40-50$ & $40-50$ & $40-50$ & $40-50$ \\ \hline
Disease Status & T2D & T2D & No Diabetes & No Diabetes & T2D & T2D \\ \hline
Medications & metformin & metformin & --- & --- & metformin & metformin, insulin pump \\ \hline
Wore Continuous Glucose Monitor & --- & --- & Yes & Yes & --- & Yes \\ \hline
Total $\#$ glucose measurements &  $80$ &  $80$ & $80$ & $80$ & $80$ & $80$ \\ \hline
Total $\#$ meals recorded &  $35$ &  $27$ & $57$ & $48$ & $41$ & $44$ \\ \hline
Total $\#$ days measured & $8$ & $10$ & $15$ & $25$ & $31$ & $32$ \\ \hline
Mean measured glucose & $111 \pm 23$  & $140 \pm 33$  & $92 \pm 16$  & $99 \pm 16$ & $112 \pm 17$  & $96 \pm 19$ \\ \hline
Training Set: $\#$ glucose measurements &  $40$ &  $40$ & $40$ & $40$ & $40$ & $40$ \\ \hline
Testing Set: $\#$ glucose measurements &  $40$ &  $40$ & $40$ & $40$ & $40$ & $40$ \\ \hline
Training Set: $\#$ meals &  $17$ &  $15$ & $27$ & $23$ & $20$ & $21$ \\ \hline
Testing Set: $\#$ meals &  $17$ &  $11$ & $29$ & $25$ & $20$ & $22$ \\ \hline
Training Set: $\#$ days measured & $4$ & $5$ & $9$ & $11$ & $5$ & $9$ \\ \hline
Testing Set: $\#$ days measured & $4$ & $5$ & $6$ & $14$ & $26$ & $23$ \\ \hline
Training Set: Mean measured glucose & $110 \pm 20$  & $138 \pm 33$  & $90 \pm 11$  & $101 \pm 13$ & $115 \pm 17$  & $102 \pm 22$ \\ \hline
Testing Set: Mean measured glucose & $111 \pm 26$  & $142 \pm 33$  & $93 \pm 20$  & $98 \pm 18$ & $109 \pm 17$  & $91 \pm 13$ \\ \hline
\end{tabular}
\begin{flushleft} Demographic information and summary statistics are reported for the six participants whose retrospectively collected data are included in the study. We note that each participant exhibited very different data collection habits, such that training and testing data (each of which contain 40 blood glucose measurements) include varying numbers of recorded meals and span different time-lengths due to individual differences in self-monitoring frequency. 
\end{flushleft}
\end{adjustwidth}
\end{table}

\begin{figure}[!h]
\begin{adjustwidth}{-2.25in}{0in} 
  \centering
  \subfloat[Typical T2D data]{\includegraphics[scale=0.2]{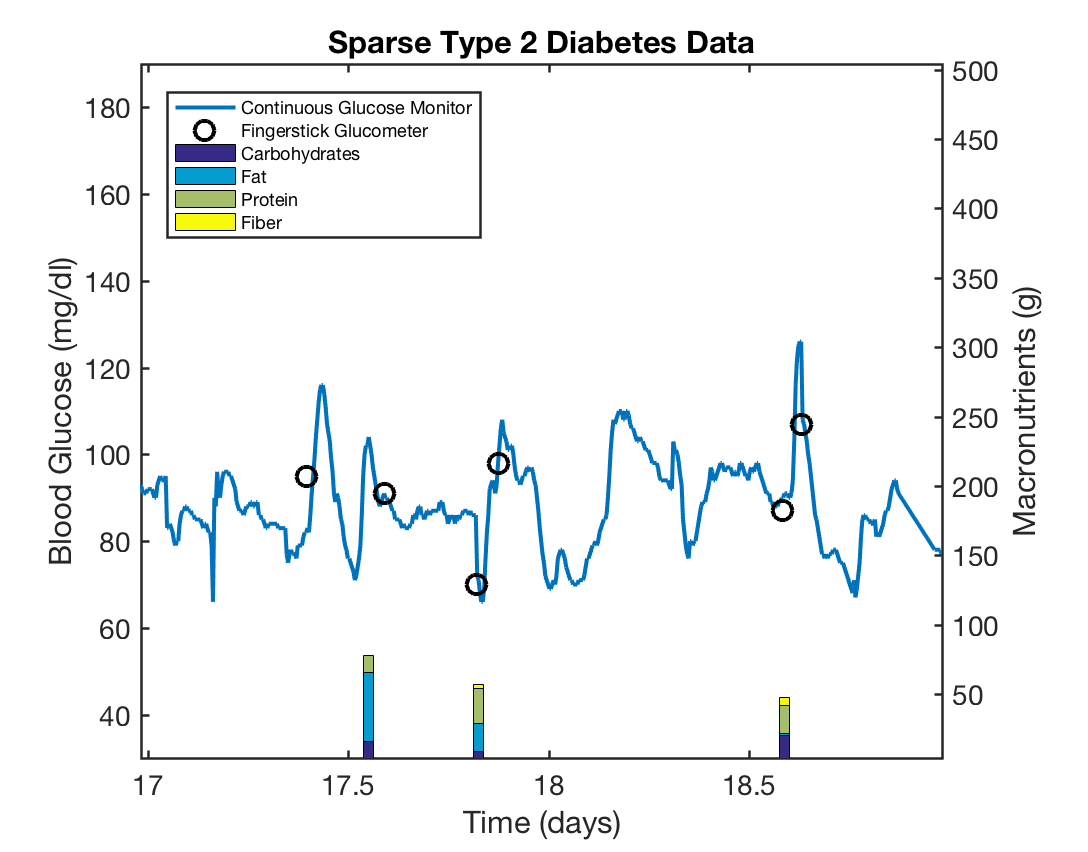}
	\label{fig:sparse_Diabetic}}
  \subfloat[High-frequency, non-T2D data]{\includegraphics[scale=0.2]{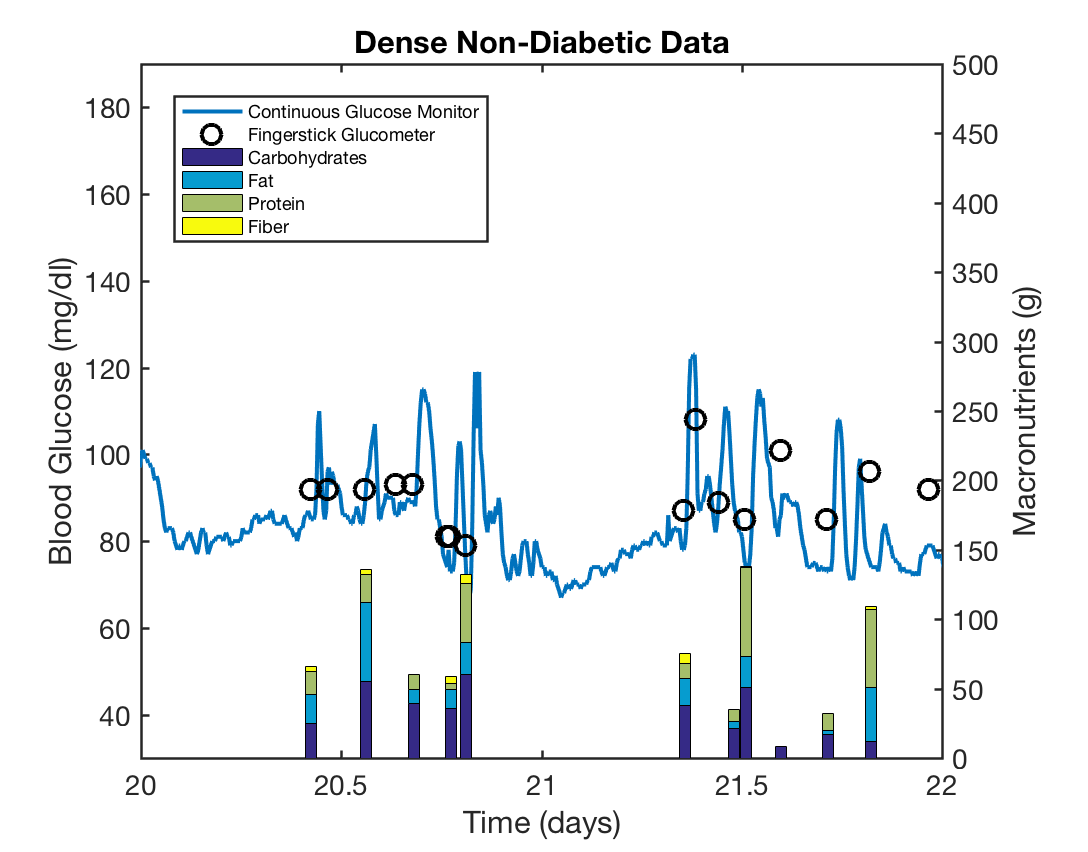}
	\label{fig:sparse_NonDiabetic}}
  \caption{{\bf Sparsity of self-monitoring data.} Here, we show qualitatively that typical diabetes self-monitoring data are sparse relative to the frequency of glucose oscillations, making apperent the challenge of reconstructing continuous dynamics from real-world data. Fingerstick measurements are maximally taken once before and after meals; but users are likely to measure less frequently and report only a subset of their actual meals.
  We overlay fingerstick measurements with passively collected glucose signals from an implantable glucose sensor, and annotate the graph with nutritional intake.
 Fig~1a is from a participant with diabetes who recorded a typical amount of self-monitoring data: 3 meals and 6 BG measurements over the course of two days. Fig~1b is from a non-diabetic participant who recorded a maximal amount of self-monitoring data: 11 meals and 15 BG measurements over the course of two days.}
  \label{fig:sparseData}
  \end{adjustwidth}
\end{figure}

Blood glucose and nutrition data were collected retrospectively from six participants, four with type 2 diabetes and two without diabetes, using custom-designed mobile applications for capturing self-monitoring data. All participants were asked to use fingerstick glucometers to record blood glucose levels once before and at least once after meals, along with a timestamp, photo, and text-based description of their meals. The nutritional content of the meals was then assessed by a team of certified diabetes educators and dietitians based on meals' photographs and textual descriptions provided by the participants. In addition to fingerstick glucose measurements, three participants also wore a subcutaneously implanted glucose monitoring device (Dexcom G4 Platinum), which samples interstitial glucose levels at 5 minute intervals. These continuous monitoring devices are not standard of care for people with type 2 diabetes; however, their high frequency measurements provide us with one type of benchmark for the ground truth of underlying glucose dynamics.

Data collected from six participants is summarized in Table~\ref{table:participants}, and includes sparse, irregularly sampled fingerstick glucose measurements, nutritional assesments of self-reported meals, and, in three participants, passively collected continuous glucose readings from an implanted sensor.
In Fig~1, we present raw blood glucose data collected by two participants (P3 and P6), and overlay these data with a glucose timeseries from the implanted continuous monitor (sampled every five minutes). We see that in both cases, the task of reconstructing the glucose dynamics from sparse measurements is highly non-trivial, and would be significantly aided by an understanding of the underlying dynamics. For this reason, we wish to incorporate validated models of the glucose-insulin system into our forecasting methodology.

Because these data were collected under the real-world constraints of chronic disease self-management, they can be sparse, noisy, and biased.  Meals can often be altered or omitted without record, and pre and post-meal glucose readings are not always reported. Blood glucose levels were entered manually into the mobile application, and thus can also be subject to not only measurement noise, but also entry-errors and delays in recording.

In summary, we acquire three types of data: 1) fingerstick blood glucose measurements taken at the discretion of each of the 6 participants (roughly 3-10 times per day), 2) measurements of interstitial glucose that are passively collected by an implanted device in 3 participants (roughly every 5 minutes), and 3) estimates of carbohydrate consumption over time (roughly 1-5 meals per day) determined by a dietitian's analysis of the daily meal logs reported by each participant.

\subsection{Experimental Design}
We wish to design an experiment that evaluates different approaches to fitting and forecasting self-monitoring data using an ultradian model of glucose dynamics. One approach, previously evaluated by Albers et al., is to apply sequential data assimilation (e.g. unscented Kalman filter) to all data. More complex approaches, however, may improve prediction performance. For example, one can optimize model parameters on a window of past data before running a filter forward. In principal, this can be done with any parameter estimation technique, and any non-linear stochastic filter. Alternatively, one can perform this same parameter optimization, and then run the model forward to forecast, \emph{without} updating states based on new measurements--in this case, forecasts are still driven by nutrition, but state-measurements in the forecasting horizon are ignored. 

In order to evaluate combinations of parameter estimation and state-forecasting methods, data from each participant was split into training and testing sections.
The final $40$ blood glucose measurements in each participant's data set were 
designated as the testing set, and the preceeding $40$ measurements were selected as the training set.
Multiple parameter estimation methods (dual unscented Kalman filter (UKF), Nelder-Mead, and Markov Chain Monte Carlo (MCMC)) were run over each participant's training set.
Then, the parameter estimates from each of these training runs were used by multiple online forecasting methods (state UKF, dual UKF, and an unfiltered model) to predict values in the testing set.

A state UKF, dual UKF, and an unfiltered model were used to sequentially forecast data in the testing set. 
The state UKF is selected for its efficacy as a prediction-correction scheme that updates unknown states given partial observations of the state vector.
The dual UKF is selected for the identical properties of the state UKF, plus its ability to track parameters for further online model improvements.
An unfiltered model is selected because it tests only the trained model and does not attempt to adjust the state space or model parameters.
Each pair of training and testing methods are evaluated for predictive performance on the test set using two standard model selection criteria and one diabetes-specific error metric that weights errors according to clinical significance.

\subsection{Data assimilation framework}
The overarching framework for the physiologic models that we consider
in this paper are dynamical models of the form
\begin{equation}
\label{eq:modelOverarching}
\frac{dx}{dt}=F(x,t;\theta).
\end{equation}

Here $x$ represents physiological state variables (e.g.
glucose and insulin concentrations) and $\theta$ represents physiological
parameters (e.g. rate constants, compartmental volumes).
We assume that the function $F$ is comprised of: 1) a time-independent component, $F_0(x,\theta)$,
that represents the interactions amongst physiological
parameters; and 2) a known time-dependent driving, $\xis(t,\theta)$, that can represent, for example,
aspects of the $24$ hour cycle and, in particular, nutritional intake. 
Thus
\begin{equation}
\label{eq:rhs}
F(x,t;\theta)=F_0(x;\theta)+\xis(t,\theta).
\end{equation}
Kalman filtering type methods may assume an additional white noise model error
to account for missing effects not included the \eqref{eq:rhs}; we will discuss
this in detail where it arises. 

We assume that data (in this case, blood glucose measurements) is available to us in the form of a set of
partial and noisy observations $\{y_k\}_{k=1}^{K}$ at times $\{t_k\}_{k=1}^K$.
Specifically we have
\begin{equation}
\label{eq:data}
y_k=H\bigl(x(t_k)\bigr)+\eta_k
\end{equation}
where $H$ picks out the subset of the physiologic state vector $x$ that can be observed;
in our case, only glucose states can be directly observed, while other states, like insulin concentrations, cannot be measured easily or safely. 
For simplicity we will model the observational
noise $\{\eta_k\}_{k=1}^{K}$ as a vector valued Gaussian random variable with mean zero and 
covariance $\Sigma_k$; typically we assume that $\Sigma_k=\Sigma_0$,
and is thus independent of the observation time.

\subsection{Ultradian model of glucose-insulin dynamics}
\begin{figure}[!h]
    \centering
    \includegraphics[scale=0.5]{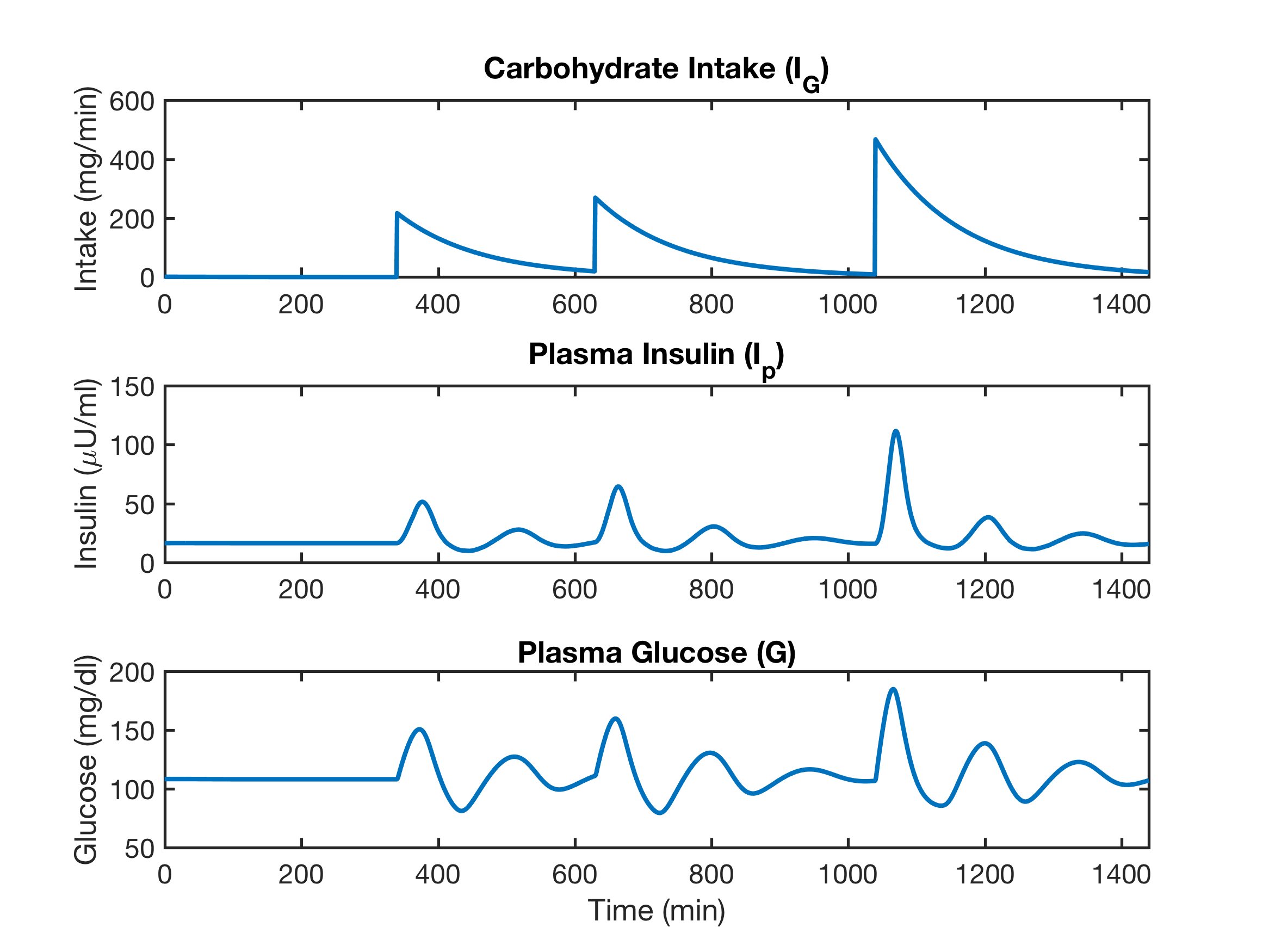}
    \caption{{\bf Ultradian model dynamics.} Here we show the oscillating dynamics of the glucose-insulin response in response to a simple exponentially decaying nutritional driver $I_G$, as governed by the model.}
    \label{fig:UltradianModel}
\end{figure}

In this paper, we outline an approach that can apply to any mechanistic model of the general form defined in Equation 1.
Here, we employ one particular model---a simple ultradian model of glucose dynamics represented by a system of six differential equations~\cite{sturisUltradian,keener_mathematical_2009}. 
The primary state variables are the glucose concentration, $G$, the plasma insulin concentration, $I_{p}$, and the interstitial insulin concentration, $I_{s}$ (a representative depiction of their evolution is shown in Fig~2); these three state variables are appended
with a three stage filter $(h_1,h_2,h_3)$ which encodes a non-linear delayed response
of the plasma insulin to glucose levels. The resulting 
ordinary differential equations have the form
\begin{subequations}
\label{eq:modelUltradian}
\begin{eqnarray}
\frac{dI_p}{dt} & = &  f_1(G)-E\bigl(\frac{I_{p}}{V_{p}}-\frac{I_i}{V_{i}}\bigr)-\frac{I_{p}}{t_{p}}\\
\frac{dI_i}{dt} & = & E\bigl(\frac{I_{p}}{V_{p}}-\frac{I_i}{V_{i}}\bigr)-\frac{I_{i}}{t_{i}}\\
\frac{dG}{dt} & = & f_4(h_3)+I_{G}(t)-f_2(G)-f_3(I_i)G\\
\frac{dh_1}{dt} & = & \frac{1}{t_d}\bigl(I_p-h_1\bigr) \\
\frac{dh_2}{dt} & = & \frac{1}{t_d}\bigl(h_1-h_2\bigr) \\
\frac{dh_3}{dt} & = & \frac{1}{t_d}\bigl(h_2-h_3\bigr)
\end{eqnarray}
\end{subequations}

The nutritional driver of the model $I_G(t)$ (shown in Fig~2) is defined as follows over $N$ discrete nutrition events, where $k$ is the decay constant and nutrition event $j$ occurs at time $t_j$ with carbohydrate quantity $m_j$~\cite{albers_personalized_2017}:
\begin{equation}
I_G(t) = \sum^N_{j=1}{\frac{m_j k}{60}\exp(k(t_j-t))} \text{    , where } N=\#\{t_j<t\}.
\end{equation}

Here,
$f_1(G)$ represents the rate of insulin production,
$f_2(G)$ represents insulin-independent glucose utilization,
$f_3(I_i)G$ represents insulin-dependent glucose utilization,
$f_4(h_3)$ represents delayed insulin-dependent glucose utilization;
the specific functional forms are taken as follows:

\begin{eqnarray}
f_1(G) & = & \frac{R_m}{1+ \exp(\frac{-G}{V_g c_1} + a_1)} \\
f_2(G) & = & U_b(1-\exp(\frac{-G}{C_2V_g})) \\
f_3(I_i) & = & \frac{1}{C_3 V_g}( U_0 + \frac{U_m - U_0}{1 + (\kappa I_i)^{-\beta}}) \\
f_4(h_3) & = & \frac{R_g}{1 + \exp(\alpha (\frac{h_3}{C_5 V_p}  -1))} \\
\kappa & = & \frac{1}{C_4} (\frac{1}{V_i} - \frac{1}{E t_i})
\end{eqnarray}

In the abstract notation we have $x=(I_p,I_i,G,h_1,h_2,h_3)^T$.
The systematic forcing to the model is given by 
$\xis(t;\theta)=(0,0,I_{G}(t),0,0,0)^T$,
and represents the external sources of glucose from nutritional intake, which we
assume to be known.
The function $F_0(x;\theta)$ describes the remainder of the vector that appears 
on the right hand-side of Eq~(\ref{eq:modelUltradian}). 
The observation operator $H(x)=G$, since we observe only glucose concentrations.

By measuring nutrition, we observe $I_{G}(t)$ and hence, $\xis(t;\theta)$, up to its decay constant $k$. We do not consider uncertainty around nutritional observation; this may be a worthwhile complexity to introduce in the future. In this way, we acquire information about the most important parts of the system--the non-autonomous driving force, $\xis(t;\theta)$ and the primary clinical response variable $G(t)$.


Here $\theta$
comprises the parameters $(V_p,V_i,E,t_p,t_i,t_d)^T$ as well as
additional parameters entering the functions $\{f_i(\cdot)\}_{i=1}^4$ and
$I_{G}$, all of which, in principal, are not known.
However, we can consider any subset of parameters to be known, and these parameters can be
removed from $\theta$ and fixed at assumed values. We always fix the plasma glucose volume parameter, $V_G$, which relates the glucose mass in the model to measured glucose concentrations, due to its obvious lack of identifiability.
Thus, $\theta$ will comprise only the parameters that we wish to estimate from the data.
We also emphasize that while we consider $\theta$ to be constant in time, $\theta$ can also be considered a vector of time-dependent functions. 

We will use the model equations \eqref{eq:modelUltradian} directly, 
to estimate parameters $\theta$ offline
and, having done so, to predict future system behavior; and we will use equations
\eqref{eq:modelUltradian} with additional white noise forcing 
as part of a prediction-assimilation cycle, a filter, 
which estimates parameters and/or states online.

\subsection{State Dynamics}
\label{sec:Unfiltered}
While we typically assume that filtered state-updates improve system tracking, this assumption must be tested. The utility of a filter is subject to, among other things, the sensitivity of its model to perturbations, and the sensitivity of a non-linear dynamical system is determined, in part, by its parameters. It is conceivable that parameter regimes exist for the ultradian glucose model, in which a filter that updates states based on new glucose measurements can actually cause degradations in predictive performance. For this reason, we evaluate the predictive performance of an unfiltered state model, in which the ultradian model is driven by nutrition, but ignores new glucose measurements and does not update its states. 

Concretely, we consider Eq~(\ref{eq:modelOverarching}) (in this case, the ultradian model) with fixed initial conditions, $x(0)$ and $\theta$, non-time-varying parameters, and no noise $\xin(t;\theta)$.
In this setup, the sole driver of the model is $I_G(t)$, which is explicitly defined by the data, given a quantity and time of carbohydrate intake. 
Because both $I_G(t)$ and individual single simulations of the model are very fast to compute (order seconds), we can produce a glucose forecast at any time $\tau$, provided that $I_G(t)$ is defined for all $t<\tau$. 
Given $x(0)$ and $\theta$, the non-autonomous differential equation 
Eq~(\ref{eq:modelOverarching}) defines a time-dependent map to the solutions at time $t$:
\begin{equation}
\label{eq:MAP}
x(t)=\Psi(x(s),t,s,\theta).
\end{equation}
With this notation it will be helpful in what follows to define
\begin{equation}
\label{eq:observe}
{\mathcal G}(\theta)=\{H\bigl(x(t_k)\bigr\}_{k=0}^{K}.
\end{equation}
This accumulates the solution at observation times, projected into the observation
space (glucose variables in our setting). The basic task at hand is to choose
$\theta$ to match this function ${\mathcal G}(\theta)$ to observations.
Methods to do this will be addressed in subsequent subsections.

We perform the computation of $\Psi$, and hence of ${\mathcal G}$, 
by using a MATLAB single-step ODE solver, \texttt{ode23}, which implements a Runge-Kutta method based on the Bogacki-Shampine (2, 3) pair.

\subsection{Filtered State Dynamics}
Here, we use a non-linear stochastic filtering methodology to track glucose-insulin state dynamics, learn a subset of each participant's physiologic parameters, and quickly issue in-the-moment forecasts. In our previous work, we employed a dual unscented Kalman filter (UKF), due to its efficiency and demonstrated success in a wide range of problems, and showed that a dual UKF was able to train consistently to different participants.

The UKF functions as a prediction-correction scheme, where forecasts are updated given each new measurement. A dual UKF runs two separate filters, where one filter updates the states given each new measurement, and the second filter updates model parameters given each new measurement and updated state. A state UKF is identical, but lacks the second parameter filter.

In this work, we employ the state and dual UKF formulations as described by Albers et al..
The model proceeeds by predicting between data aquisition times $t_k$, using \eqref{eq:MAP}, and then incorporating data using Bayes' theorem with approximate Gaussian statistics. 
If $x_k$ is the state at time $t_k$ after aquisition of data $y_k$, then
the predicted state is a continuous time variable given by
\begin{equation}
\label{predict}
\hat{x}(t)=\Psi(x_k,t,t_k,\theta), \quad t \in (t_k,t)
\end{equation}
and the forecast at the next measurement time $t_{k+1}$
is then corrected to agree with measurements $y_k$. This is done
via a Kalman update of the form 
\begin{equation}
x_{k+1}= \hat{x}(t_{k+1}) + K_k\Bigl(y_k - H\bigl(\hat{x}(t_{k+1})\bigr)\Bigr).
\end{equation}
The Kalman gain $K_k$ is derived from a linearization of the dynamics of the flow about
$\hat{x}(t).$
The prediction step is also sometimes appended with random noise $\xin;$ the variance
of this noise enters the Kalman gain.
The method we use is termed the state UKF.

This methodology may be extended to learn the parameters as well, based on an
assumed dynamics of the form
$$\frac{d\theta}{dt}=0,$$
leading to the dual unscented Kalman filter.
Here, a dual UKF is used as one of the parameter estimation schemes during the training phase, and is used to learn parameters ($C_1$, $C_3$, $U_b$).
Both dual and state UKFs are run during the testing phase, and are evaluated for forecasting performance. UKF hyperparameters are not tuned, and are fixed at the values reported by Albers et al..

\subsection{Nelder-Mead Optimization}

\begin{algorithm}
\caption{Nelder-Mead Simplex Optimization}\label{NelderMead}
\label{alg:NelderMead}
\begin{algorithmic}[1]
\Procedure{Nelder-Mead Simplex Optimization}{$y$} 
	\State $\rho = 1$
	\State $\chi = 2$
	\State $\gamma = 0.5$
	\State $\sigma = 0.5$

	\For{$i=1:n+1$}
		\State Choose $\theta_i \thicksim \bbP(\theta)$  \Comment{Draw random theta initialization from prior}
		\State Compute ${\cal G}(\theta_i)$ according to Eqs~(\ref{eq:MAP}, \ref{eq:observe}) \Comment{Use an ode solver, like \texttt{ode23}}
		\State Compute $\Phi(\theta_i;y)$ according to Eq~(\ref{eq:NMmisfit}) \Comment{Misfit of ${\cal G}(\theta_i)$}
	\EndFor
	\For {k=1:N}
		\State Order $\{\theta_i\}$ such that $\Phi(\theta_1;y) \le \Phi(\theta_2;y) \le ... \le \Phi(\theta_{n+1};y)$ 

		\State $\theta_0 \gets \frac1n{\sum_{i=1}^{n}{\theta_i}}$ \Comment{Compute centroid}
		\State $\theta_r \gets \theta_0 + \rho(\theta_0 - \theta_{n+1})$ \Comment{Perform reflection}
		\State Compute $\Phi(\theta_r;y)$ \Comment{Misfit of reflected point}

		\If {$\Phi(\theta_1;y) \le \Phi(\theta_r;y) < \Phi(\theta_n;y) $}
			\State $\theta_{n+1} \gets \theta_r$ \Comment{Keep reflected point}
		\ElsIf {$\Phi(\theta_r;y) < \Phi(\theta_1;y)$}
			\State $\theta_e \gets \theta_0 + \chi(\theta_r - \theta_0)$ \Comment{Perform expansion}
			\State Compute $\Phi(\theta_e;y)$ \Comment{Misfit of expanded point}
			\If {$\Phi(\theta_e;y) < \Phi(\theta_r;y)$}
				\State $\theta_{n+1} \gets \theta_e$ \Comment{Keep expanded point}
			\Else
				\State $\theta_{n+1} \gets \theta_r$ \Comment{Keep reflected point}
			\EndIf
		\Else \Comment{ Here $\Phi(\theta_r;y) \ge \Phi(\theta_n;y)$}
			\If {$\Phi(\theta_r;y) < \Phi(\theta_{n+1};y)$}
				\State $\theta_c \gets \theta_0 + \gamma(\theta_r - \theta_0)$ \Comment{Perform outside contraction}
			\Else
				\State $\theta_c \gets \theta_0 + \gamma(\theta_{n+1} - \theta_0)$ \Comment{Perform inside contraction}
			\EndIf
			\State Compute $\Phi(\theta_c;y)$ \Comment{Misfit of contracted point}
			\If {$\Phi(\theta_c;y) < \Phi(\theta_{n+1};y)$}
				\State $\theta_{n+1} \gets \theta_c$ \Comment{Keep contracted point}
			\Else 
				\State Shrink all points except best ($\theta_1$)
				\For{$i=2:n+1$}
					\State $\theta_i \gets \theta_1 + \sigma(\theta_i - \theta_1)$
				\EndFor
			\EndIf
		\EndIf
	\EndFor
\EndProcedure
\end{algorithmic}
\end{algorithm}

In order to evaluate parameter optimization over training data, we test the downhill simplex optimization routine (often referred to as Nelder-Mead optimization). 
Nelder-Mead optimization is a popular derivative-free approach for constrained non-linear minimization problems, and is often used in parameter estimation for biological models by finding parameter values that minimize a cost function describing the misfit between data and model forecasts. Here, we define our cost function as the sum of squared errors between measurements and predictions (equivalent to MSE in Eq~(\ref{eq:MSE})), and run the ultradian model with fixed initial conditions and a fully specified $I_G(t)$. 

The optimization algorithm is specified in Alg~\ref{alg:NelderMead}, which relies on the following definitions. We let ${\cal G}(\theta)$ define the forward mapping of the parameter vector $\theta$ and all other assumptions of the model. Here, we recall the definition
of ${\cal G}(\theta)$ from Eq~(\ref{eq:observe}). 
The function that we wish to minimize is the misfit between data and forecasts, which we refer to as $\Phi(\theta;y)$:

\begin{equation}
\label{eq:NMmisfit}
\Phi(\theta;y)=\Bigl\|\bigl(y-{\mathcal G}(\theta)\bigr)\Bigr\|^2.
\end{equation}

We implemented a constrained version of the algorithm, which is found in MATLAB central as \texttt{fminsearchbnd}~\cite{fminsearchbnd}, which layers a simple transformation of variables on top of the standard \texttt{fminsearch} Matlab function. 
Model parameters were bounded from below at small non-negative values to avoid non-physical representations. We experimentally determined lower bounds for parameters that appear as denominators in order to avoid issues of tractability. Parameters were not bounded from above.

\subsection{Markov Chain Monte Carlo methods}
In order to evaluate Bayesian parameter estimation over training data, we employ a Metropolis-within-Gibbs, random-walk, Markov Chain Monte Carlo (MCMC) routine. 

Begin with Bayes theorem:
\begin{align}
p(\theta | y) = k p (y | \theta) p(\theta) \\
p(\theta | y)= \mathcal{L}(\theta, y) p(\theta) 
\end{align}
where $p(\theta | y)$ is the posterior, or the model predictions given data, $y$ are data, $\theta$ are parameters, $\mathcal{L}$ is the likelihood function, and $k$ is a normalizing constant depending on the data. 

We place a prior probability distribution
$\bbP(\theta)$ on the parameters $\theta$, reflecting our beliefs
about them before data are obtained from a particular patient.
The equations \eqref{eq:modelOverarching} and \eqref{eq:data} define
a mapping from the parameters $\theta$ 
into the data $y=(y_1^T,\cdots, y_K^T)^T.$ We thus
have
\begin{equation}
\label{eq:smooth}
y={\cal G}(\theta)+\eta
\end{equation}
where $\eta=(\eta_1^T,\cdots, \eta_K^T)^T$ is the noise in the set of observations, which
we define as a Gaussian with mean zero and covariance $\Sigma$. ${\cal G}(\theta)$ defines
the forward mapping of the vector $\theta$ and all other assumptions of the model. Here, we draw on
the definitions from \eqref{eq:modelOverarching} and \eqref{eq:data} to write
\begin{equation}
\label{eq:G}
{\cal G}(\theta)=\{H(\int_{0}^{t_k}{F_0(\hat{x}(t),\theta) + \xis(t,\theta) dt})\}_{k=1}^K
\end{equation}
This defines a likelihood for the data $\bbP(y|\theta).$
Bayes' rule then gives us an equation (upto normalization) for the
posterior probability on the parameters given the data, $\bbP(\theta|y)$: 
\begin{equation}
\label{eq:bayes}
\bbP(\theta|y) \propto \bbP(y|\theta)\bbP(\theta).
\end{equation}
The equation \eqref{eq:smooth} shows that $y|\theta$ is a Gaussian
with mean ${\cal G}(\theta)$ and covariance $\Sigma$.
Thus the probability density function $\bbP(y|\theta)$ is proportional to
$\exp\bigl(-\Phi(\theta;y)\bigr)$
where
\begin{equation}
\label{eq:misfit}
\Phi(\theta;y)=\frac{1}{2}\Bigl\|\Sigma^{-\frac12}\bigl(y-{\mathcal G}(\theta)\bigr)\Bigr\|^2.
\end{equation}
For simplicity, we will assume that the prior on $\theta$ is Uniform over a set $S$ to allow us to easily bound parameters from below. Specifically, we define positive lower bounds for each element of $\theta$ by experimentally determining the smallest values that were still computationally tractable (without worse than a ten-fold slow-down of computation) in \eqref{eq:MAP}. We impose no upper bound on $\theta$.
We then have that $\bbP(\theta)$ is proportional to $\Phi_0(\theta)$ where
\begin{equation}
\label{eq:misfit}
  \Phi_0(\theta) =
  \begin{cases}
       0 & \text{if $\theta \notin S$} \\
       1 & \text{if $\theta \in S$} \\
  \end{cases}
\end{equation}

Thus $\bbP(\theta|y)$ is proportional to $\Phi_0(\theta)\exp\bigl(-\Phi(\theta;y)\bigr).$
Although the prior and likelihood are constructed through a uniform distribution on
$\theta$ and a Gaussian on $y|\theta$, the posterior probability on $\theta|y$
is not Gaussian unless the map ${\mathcal G}$ is linear, which it will not
be for most of our applications. As a consequence, it is necessary to use
statistical sampling methods to probe the posterior distribution
on $\theta|y$ and, in this paper, we will concentrate on the use of
Monte Carlo Markov chain (MCMC) methods. These methods produce
a Markov chain $\theta^{(k)}$ whose empirical statistics (such as
mean and covariance) approximate the desired posterior distribution
on parameters $\theta$ given data $y$. 

\begin{algorithm}
\caption{Random-Walk Metropolis-within-Gibbs MCMC}
\label{alg:MCMC}
\begin{algorithmic}[1]
\Procedure{Random-Walk Metropolis-within-Gibbs MCMC}{$y$,$C$} 
	\For{$c\gets 1:s$} \Comment{Run $s$ chains (optionally, in parallel)}
		\State Choose $\theta_1 \thicksim \bbP(\theta)$  \Comment{Draw random theta initialization from prior}
		\For{$k=1:n$} \Comment{Run $n$ iterations of MCMC}
			\For{$j=1:d$} \Comment{Loop over $d$ dimensions of $\theta_k$}
				\State $v_k^{i \not= j} \gets \theta_k^i$
				\State $v_k^{i=j} \gets \theta_k^i + \texttt{randomNormal}(0,C_{j,j})$ 	\Comment{Gaussian proposal}
				\While{$ \Phi_0(v_k)$ == 0} \Comment{Reject proposal not in prior}
					\State $v_k^{i=j} \gets \theta_k^i + \texttt{randomNormal}(0,C_{j,j})$
				\EndWhile

				\State Compute ${\cal G}(\theta)$ according to \eqref{eq:G} \Comment{Use an ode solver, like \texttt{ode23}}
				\State Compute $\Phi(v_k;y)$ according to \eqref{eq:misfit} \Comment{Misfit of proposal $v_k$}
				\State Compute $\Phi(\theta_k;y)$ according to \eqref{eq:misfit} \Comment{Misfit of ${\cal G}(\theta_k)$}
				\State $a \gets \min\{1,\exp\bigl(\Phi(\theta_k;y)-\Phi(v_k;y)\bigr)\}$

				\State $r \gets \texttt{randomUniform}(0,1)$
				\If {$a > r$}
					\State $\theta_{k+1}^j \gets v_k^j$ \Comment{Accept proposal}
				\Else
					\State $\theta_{k+1}^j \gets \theta_{k}^j$ \Comment{Reject proposal}
				\EndIf
			\EndFor
		\EndFor
	\EndFor
\EndProcedure
\end{algorithmic}
\end{algorithm}

In this work, we utilize a Random-Walk (Gaussian proposal) Metropolis-within-Gibbs approach with a uniform prior. A uniform prior is selected to facilitate placing lower bounds on parameters that cause the ODE system to become computationally intractable as they approach zero. Metropolis-within-Gibbs (sometimes refered to as "variable-at-a-time"), which iterates through proposals in each parameter dimension sequentially (as opposed to jointly), allows for higher proposal acceptance probability. While this approach can be more efficient than issuing multi-dimensional proposals, it is suboptimal when variables are highly correlated.
The algorithm employed in this work is specified in Alg~\ref{alg:MCMC}; details on this method can be found in the book by Gamerman et al.~\cite{gamerman2006markov}.

We allow 5 independent instantiations of MCMC (Alg~\ref{alg:MCMC}) to run in parallel for no more than 12 hours, and adjusted the proposal covariance, C, once over an unused subset of data from P3 to achieve acceptance rates of each variable between 20-30\%. We leverage these 5 approximations of the posterior distributions in the following ways, and treat each as a separate offline training method in this work: 1) $\mathbb{E}[\theta_k]$ (chain mean) for each MCMC chain, 2) $\argmax(\rho({\cal G}(\theta),y))$ for all $\theta_k$ across the 5 independent chains (the parameter estimate that optimized linear correlation between ${\cal G}(\theta)$ and the data $y$, defined in Eq~(\ref{eq:correlation})), and 3) $\argmin(MSE({\cal G}(\theta),y))$ for all $\theta_k$ across the 5 independent chains (the parameter estimate that optimized mean squared error between ${\cal G}(\theta)$ and the data $y$, defined in Eq~(\ref{eq:MSE})).

\subsection{Prediction performance evaluation}
We wish to compare the forecasting accuracy over the testing set achieved by different combinations of parameter estimation and forecasting methodologies. In order to do this, we must define evaluation metrics.
We use two common model selection criteria, Mean Squared Error and linear correlation, as well as a diabetes-specific metric (Parkes error grid) designed to weight forecasting errors by clinical severity. 

If $\hat{y}$ is a vector of $n$ predictions, and $y$ is the vector of observed values, then we define mean squared error (MSE) and linear correlation ($\rho$) in the typical ways:

\begin{equation}
\label{eq:MSE}
\mathrm{MSE} = \frac{1}{n}\sum{(\hat{y}_i - y_i)^2},
\end{equation}

\begin{equation}
\label{eq:correlation}
\rho = \frac{\mathrm{Cov}(\hat{y},y)}{\sigma_{\hat{y}} \sigma_{y}}.
\end{equation}

We also wish to quantify how well each method performs for an individual relative to $K$ other methods. Thus, we define a $\%$optimal criteria for the $j$th method, and report the mean of these quantities ($\%$optimal-$\mathrm{MSE}_j$ and $\%$optimal-$\rho_j$) over all participants to describe the average relative performance of the $j$th method: \\
\begin{equation}
\label{eq:fracMSE}
\% \mathrm{optimal MSE}_j = 100\%*{\frac{\min_{k \in K} \mathrm{MSE}_{k}}{\mathrm{MSE}_{j}}}
\end{equation}

\begin{equation}
\label{eq:fracCorrelation}
\% \mathrm{optimal }\rho_j = 100\%*{\frac{\mathrm{\rho}_j}{\max_{k \in K} \mathrm{\rho}_{k}}}
\end{equation}

\begin{figure}[!h]
    \centering
    \includegraphics[scale=0.3]{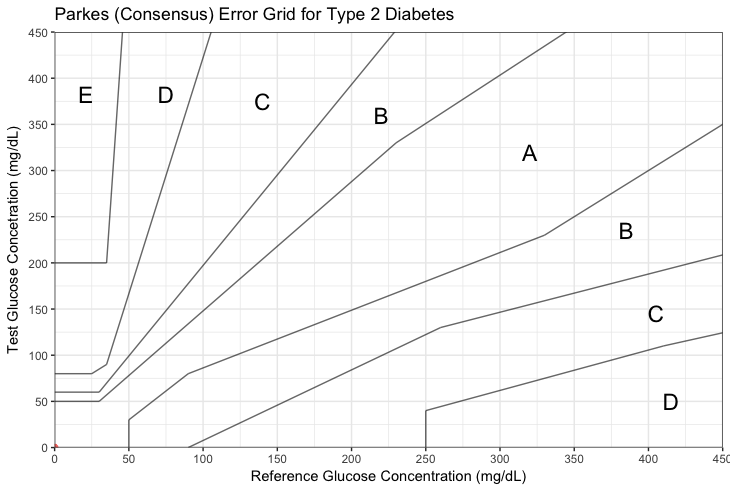}
    \caption{{\bf Parkes error zones.} The error zones for type 2 diabetes are shown to provide an intuitive understanding for regions of higher and lower clinical significance. This plot was generated using \emph{ega} package in R.}
    \label{fig:parkesGrid}
\end{figure}

We also wish to understand the clinical impact of forecasting errors, and recognize that clinical impact is not as simple as MSE or correlation.
For this reason, we employ an error classification metric created by Parkes et al. that weights errors according to clinical importance. The approach was designed to evaluate new glucometers with respect to reference glucose measurements. The authors defined five risk classifications (A: "clinically accurate measurements, no effect on clinical action"; B: "altered clinical action, little or no effect on clinical outcome"; C: "altered clinical action, likely to affect clinical outcome"; D: "altered clinical action, could have signifcant clinical risk"; E: "altered clinical action, could have dangerous consequences" and asked 100 physicians to assign a risk score to measurement-reference pairings that might be given to insulin-dependent patients with T2DM. From this, they determined the boundaries of the five severity zones, which can be depicted in the Parkes Consensus Error Grid (Fig~3). We consider any method that generates forecasts in the D or E ranges to be unacceptable, and we wish to minimize forecasts in range C and B while maximizing forecasts in range A.

\section{Results}

We predict blood glucose measurements, given sparse and noisy observations of nutrition and glucose from self-monitoring data. We wish to compare the predictive performance between approaches that leverage offline and online data assimilation. To achieve this comparison, we split each dat set into sequential training and testing sets, and evaluate pairings of training and testing methods based on predictive performance over the testing set (as measured by Mean Squared Error (MSE), linear correlation, and Parkes quality (a diabetes-specific metric that weights forecasting errors according to their clinical severity).

\subsection{Comparing offline and online approaches}
\label{smallExperiment}
As a first, broad inspection into the types of methods that tend to perform best, we select a single offline method to compare with our previously published online approach. We evaluate the following three scenarios across all six participants: 1) \emph{offline}: we use a bounded Nelder-Mead optimization to estimate parameters of the ultradian model over the training set, then use these estimates to make unfiltered glucose forecasts of the testing set 2) \emph{online}: we run a dual UKF sequentially through the training and testing sets (the approach previously applied by Albers et al.), and 3) \emph{nominal}: skip training (use parameters provided by Sturis et al.) and test with no filtering. 

\subsubsection{Performance evaluation}

\begin{table}[!ht]
\begin{adjustwidth}{-2.25in}{0in} 
\caption{{\bf Offline vs Online Comparison}}
\label{table:simpleOffline}
\begin{tabular}{|c|c|c|c|c|c|c|c|c|c|}
\hline
 & \multicolumn{3}{|c|}{MSE} & \multicolumn{3}{|c|}{Linear Correlation} & \multicolumn{3}{|c|}{Parkes Quality} \\
 \hline
Participant & Offline & Online & Nominal & Offline & Online & Nominal & Offline & Online & Nominal\\ 
\hline
P1 & \textbf{471} & 973 & 796 & \textbf{0.57} & 0.28 & -0.00 & 77.5 & 67.5 & \textbf{80.0}\\ \hline
P2 & \textbf{1204} & 1571 & 2331 & 0.05 & \textbf{0.08} & -0.28 & \textbf{87.5} & 77.5 & 42.5 \\ \hline
P3 & 289 & \textbf{234} & 536 & 0.67 & \textbf{0.71} & 0.46 & 85 & \textbf{90.0} & 87.5\\ \hline
P4 & \textbf{329} & 533 & 785 & \textbf{0.20} & 0.03 & -0.01 & \textbf{92.5} & 77.5 & 80.0\\ \hline
P5 & 924 & 605 & \textbf{539} & 0.11 & \textbf{0.36} & 0.01 & 67.5 & 62.5 & \textbf{87.5}\\ \hline
P6 & \textbf{213} & 352 & 561 & \textbf{0.32} & 0.00 & -0.18 & \textbf{92.5} & 85.0 & 87.5\\ \hline
\end{tabular} \\
\begin{flushleft} Each method (offline, online, and nominal) performed best among a subset of participants for a subset of evaluation metrics, with the offline method performing best in a majority of participants. We report results from an experiment to compare offline and online training methods over self-monitoring data-sets from people with and without diabetes, and plot results for three metrics: Mean Squared Error (MSE), linear correlation, and Parkes Quality (specifically, the percentage of forecasts with clinically insignificant errors). We compare three methods: 1) offline (Nelder-Mead parameter estimation over the training set, unfiltered forecasts over the testing set), 2) online (dual UKF over the training and testing set), and 3) nominal (untrained model generating unfiltered forecasts). We find that: 1) using the original, untrained model without any fitting or filtering is insufficient and can produce errors that negatively affect clinical outcomes (e.g. P2), 2) offline parameter estimation can provide significant improvements over an online approach (P1, P2, P4, P6), and 3) the online dual UKF method outperforms the offline approach for two of the six participants (P3, P5). 
\end{flushleft}
\end{adjustwidth}
\end{table}

\begin{figure}[!tbp]
  \centering
  \subfloat[P2]{\includegraphics[scale=0.3]{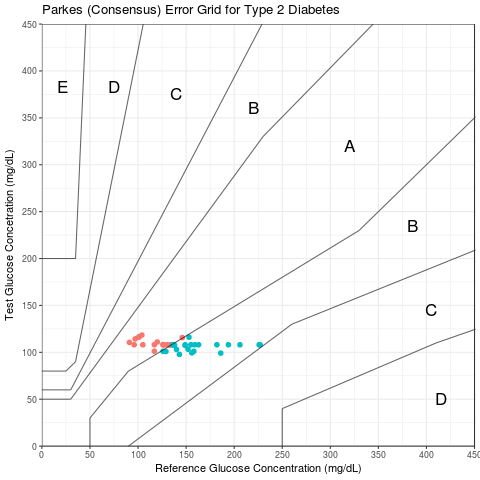}
\label{fig:ErrorGridP2}}
  \subfloat[P5]{\includegraphics[scale=0.3]{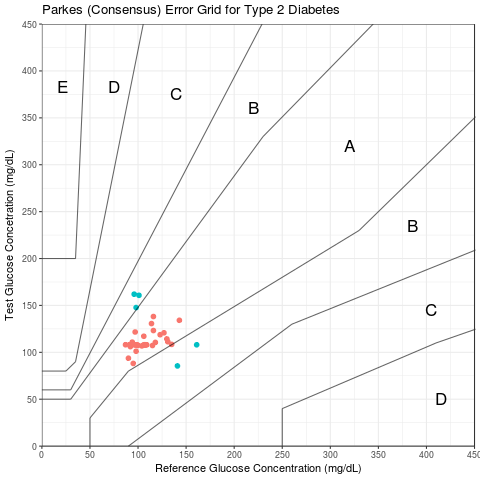}
\label{fig:ErrorGridP6}}
  \caption{The untrained ultradian glucose model can produce unexpectedly biased forecasts with clinically significant errors, depending on the patient. We plot the errors made by the untrained, unfiltered model for two participants with diabetes (P2 and P5), and plot the glucose forecast--measurement pairs on the Parkes Error Grid (the x-axis represents measurements, and the y-axis represents forecasts). Forecasts for P2 using the untrained, unfiltered model primarily occupy the B region of the error grid (58\% of forecasts fell in zone B, which signifies "altered clinical action"). Forecasts for P5, however, fell in zone A 88\% of the time, indicating that the forecasts were primarily on the level of "clinically accurate measurements".}
\label{fig:ErrorGrids}
\end{figure}

Table~\ref{table:simpleOffline} shows the predictive performance (according to MSE, linear correlation, and Parkes quality metrics) of these three methods across participants.
First, we observe that the offline method creates substantial improvement over other methods for four (P1, P2, P4, and P6) of the six participants. For these four participants, MSE is always minimized by the offline method, and linear correlation and Parkes quality are optimized in the majority of the cases through the offline method.

Second, we see that the offline method did \emph{not} present uniform advantages. For example, the data from P3 (a non-diabetic) is forecasted best using the online dual UKF approach, according to all metrics. In addition, the online method appears to be substantially better than the offline method for P5, when considering MSE and linear correlation. 

Third, we observe that the nominal model is deemed insufficient by the Parkes quality metric in one of the six participants, as it exhibits clinically significant errors more often than non-significant errors. Fig~4 shows that the majority of forecasts for P2 by the nominal model land in error region B, and approach the C-region, whose errors are deemed "likely to affect clinical outcome." Without model personalization, the severity of forecast errors scales with dissimilarity between the patient and the nominal model.

Overall, the results in Table~\ref{table:simpleOffline} demonstrate that: 1) using the original, untrained model without any fitting or filtering is insufficient and can produce clinically problematic errors (Fig~4), 2) a straight-forward approach to offline parameter estimation (i.e. Nelder-Mead) often provides significant improvements over a dual UKF online approach, however, 3) the dual UKF retains advantages over the proposed offline method for a subset of participants. The comparison also demonstrates that offline methods may remove the need for a constant stream of measurements, upon which the filtering approach, by contrast, relies heavily.
	
\subsubsection{Pointwise prediction errors}

\begin{figure}[!h]
  \centering
  \subfloat[P6]{\includegraphics[scale=0.3]{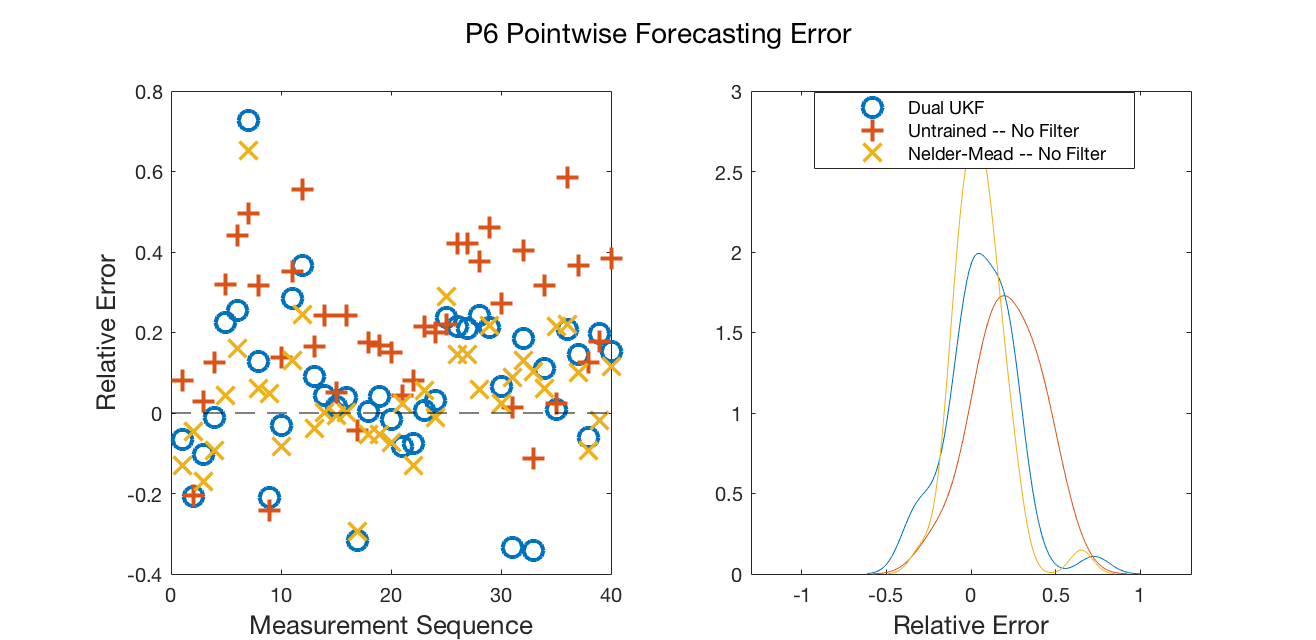}
  \label{fig:pointwiseErrorP6}}
  \par
  \subfloat[P5]{\includegraphics[scale=0.3]{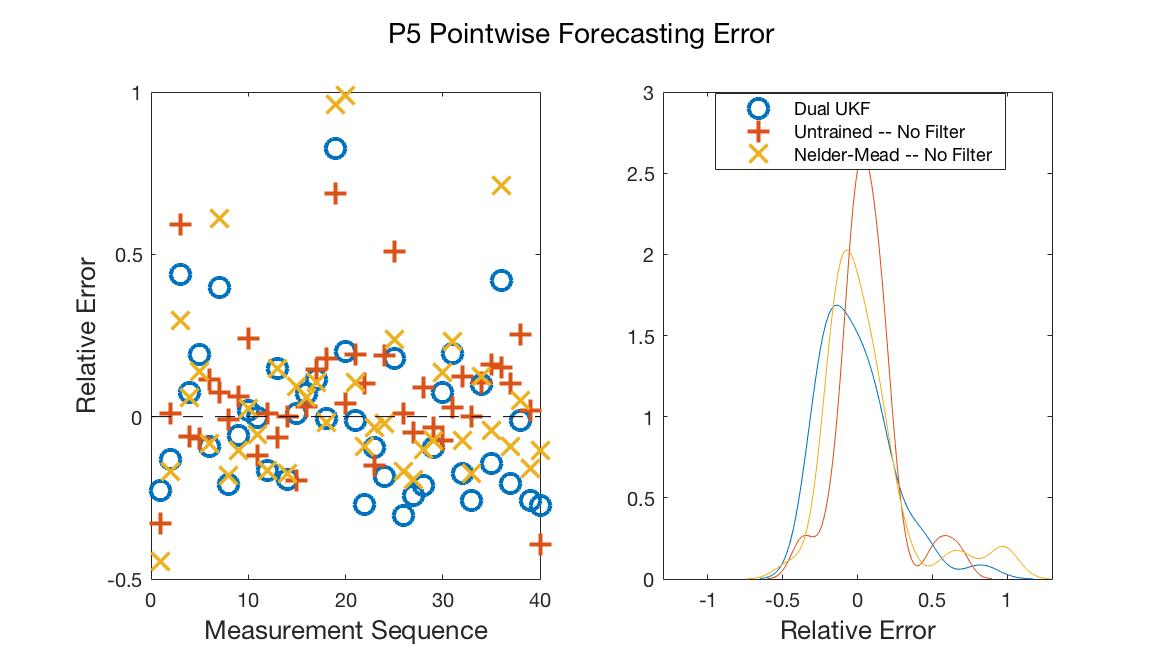}
  \label{fig:pointwiseErrorP5}}
  \caption{{\bf Pointwise Errors in offline and online forecasting methods.}
  The simple offline approach to training and forecasting differed noticeably, at times, across participants. We plot the sequence of relative measurement errors over the testing sets for P5 and P6 in Fig~5a-5b, respectively. Kernel density estimates of each of these error sequences for the three testing methods (Untrained and unfiltered, dual UKF, and Nelder-Mead-trained without filtering) are also shown. It is evident that untrained model performs worse than other methods for P6, but works surprisingly well on the P5 data.}
  \label{fig:pointwiseError}
\end{figure}

To gain an intuition for the observed forecasting errors, we examine the pointwise forecasting errors for the three experiments (online dual UKF, offline Nelder-Mead unfiltered, and nominal unfiltered model) for the P5 and P6 data set (which exhibited very different comparative performances in offline, online, and nominal settings). The left frames of Fig~5 depict the sequence of errors, and the right frames plot kernel density estimates of relative forecasting errors. 

The error sequences in the left of Fig~5 provide a qualitative description for when extreme errors occur, and which methods tend to have larger relativer error. In both participants, we observe that outliers in forecasting errors tend to be correlated across methods, indicating that the worst predictions are probably not the fault of the estimation schemes, but rather related to model insufficiences and/or data quality issues.

The density estimates on the right of Fig~5 provide a picture of the aggregate statistics of relative forecasting errors, which gives insight into the bias and variance of errors made by different methods. Evaluating the errors in this way allows for targeted mathematical approaches to error reduction.

The untrained, unfiltered forecast error distributions are right-shifted for both participants, and indeed, this method exhibited a substantial positive bias ($21\%$ relative error for P6, and $6\%$ for P5). Both the dual UKF and unfiltered Nelder-Mead forecasts are noticeably less biased ($6\%$ and $5\%$ mean relative error, respectively, in P6). In fact, UKF-based forecasts for P5 show very little bias (less than $1\%$ mean relative error). 

Interestingly, we observed that the variance in relative error is lowest in P6 when using Nelder-Mead, yet is highest in P5 with Nelder-Mead. This provides further support for our observation from Table~\ref{table:simpleOffline} that implementation of Nelder-Mead is not always a fruitful strategy.


\subsubsection{Qualitative assessment of inferred continuous dynamics}

\begin{figure}[!h]
  \centering
  \begin{adjustwidth}{-2.25in}{0in} 
  \subfloat[Untrained, No Filter]{\includegraphics[scale=0.15]{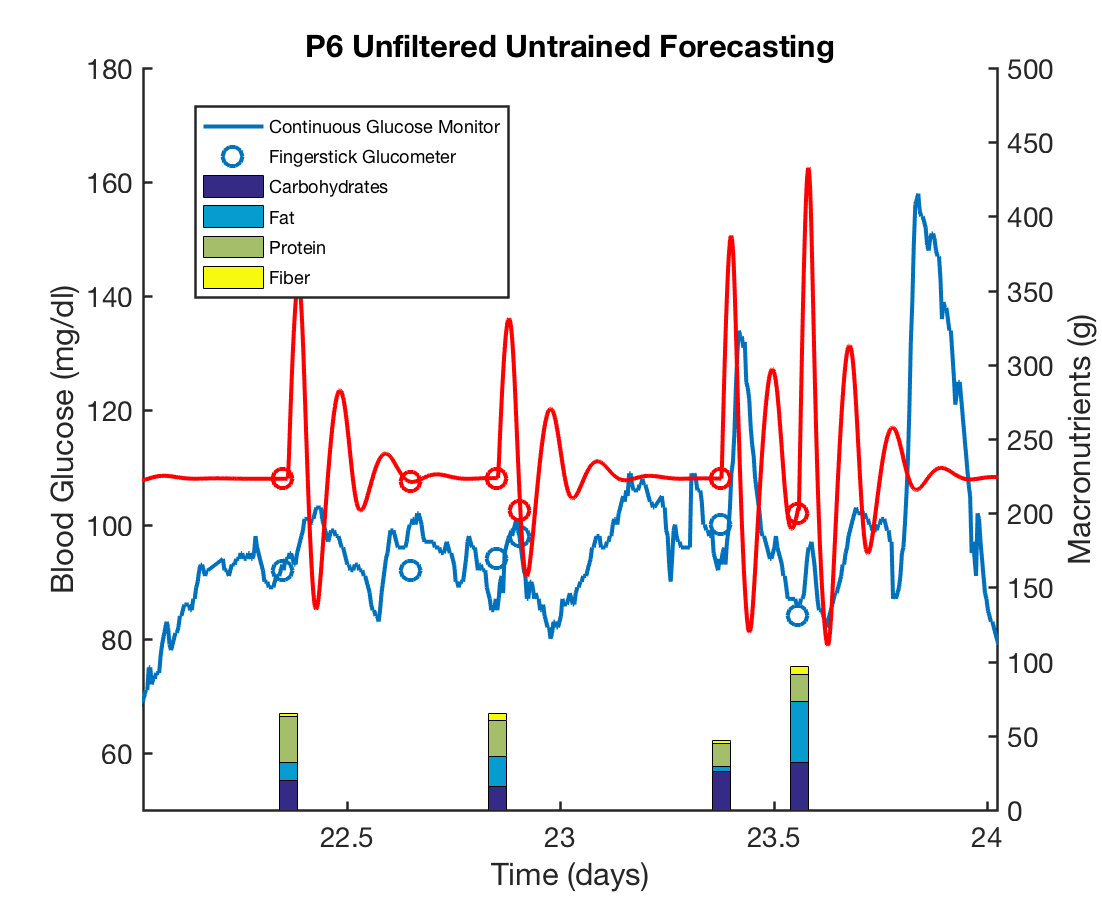}
\label{fig:curveUntrained}}
  \subfloat[Dual UKF]{\includegraphics[scale=0.15]{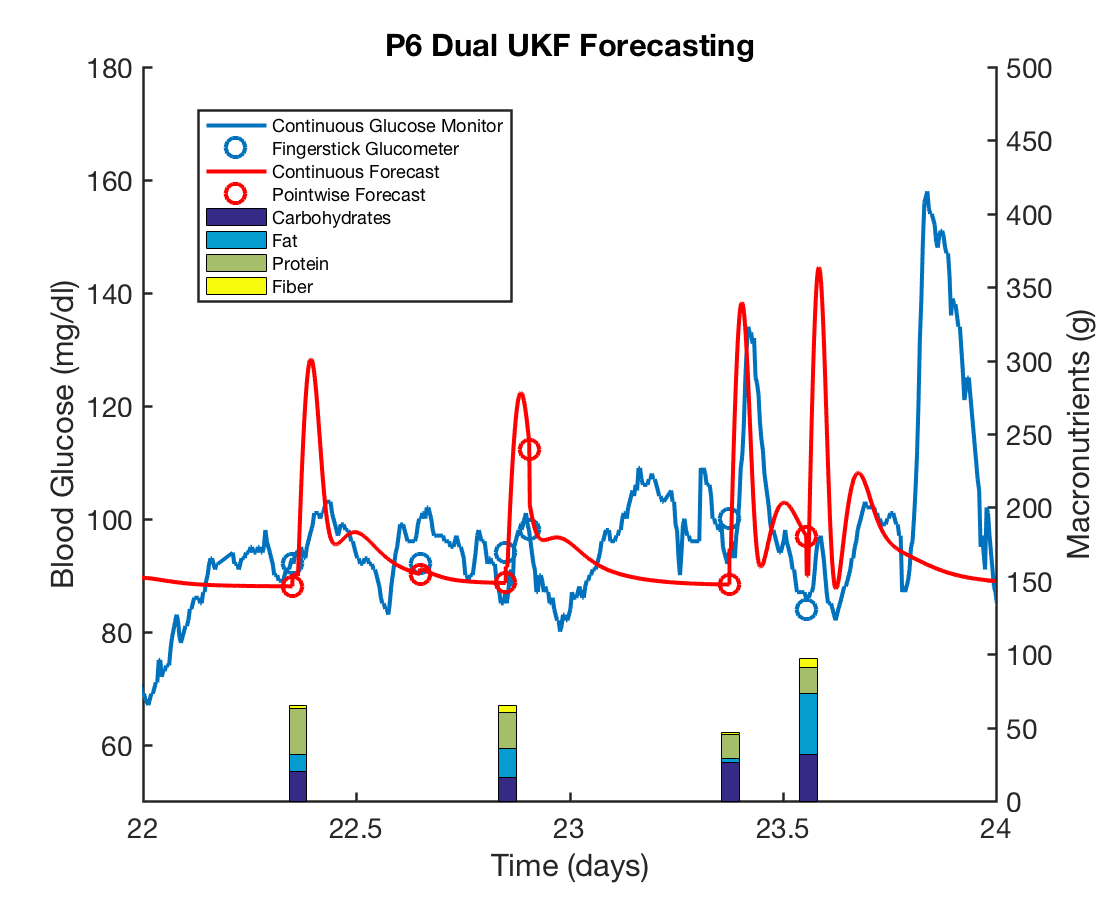}
\label{fig:curveUKF}}
  \subfloat[Nelder-Mead Trained, No Filter]{\includegraphics[scale=0.15]{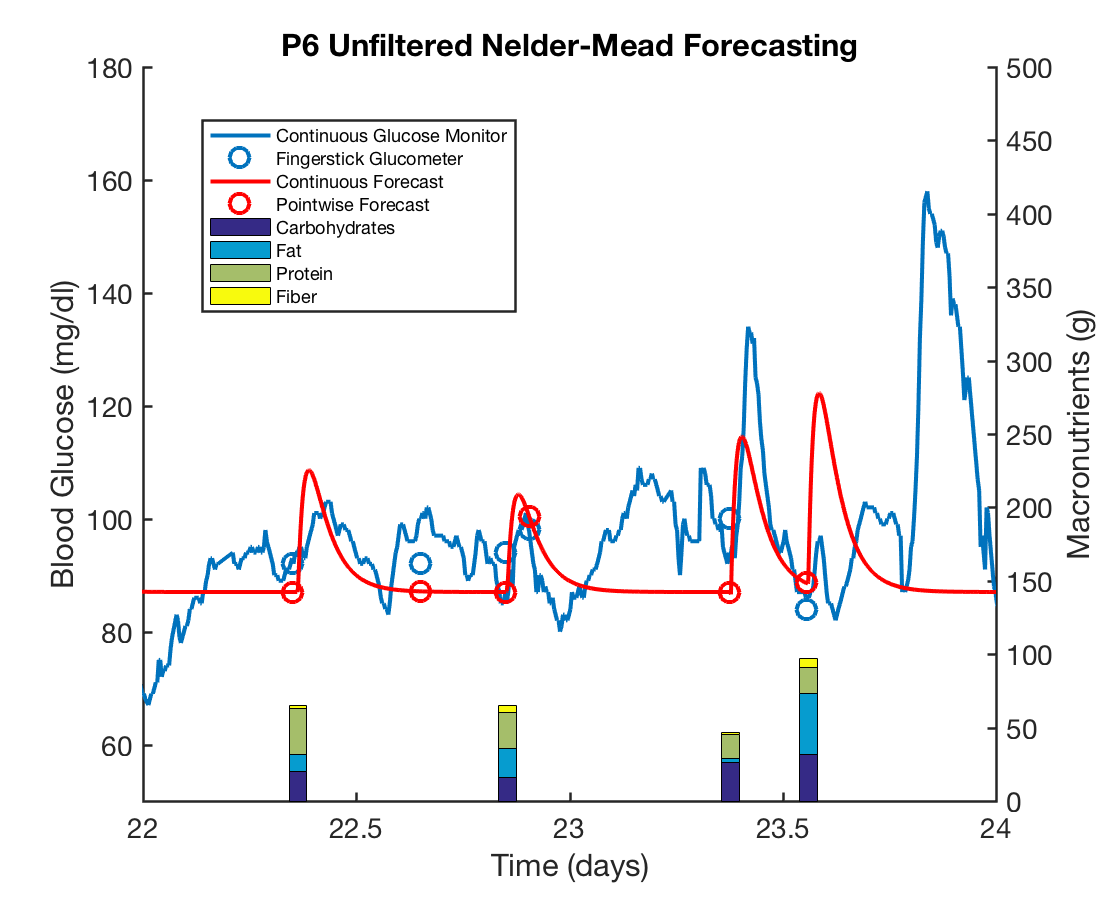}
\label{fig:curveNelder}}
  \caption{{\bf Inference of continuous dynamics.}
  Different approaches to parameter estimation of the ultradian glucose model substantially alter the dynamical properties of the inferred glucose trajectories, and represent different aspects of the dynamics measured by an implanted sensor. We plot in red the continuous glucose inferences made by the three evaluated forecasting schemes (untrained and unfiltered, dual ukf (online), and Nelder-Mead-trained unfiltered (offline)) for P6. The blue curves show measurements taken every 5 minutes from an implanted glucose sensor. Circles indicate the sparse fingerstick measurements and their corresponding point-wise forecasts. We observe that dual UKF learns to better approximate equilibrium values than the untrained model by adjusting relevant parameters. While the dual UKF retains similar oscillation dynamics as the untrained model, the offline fit model exhibits noticeably different meal responses, with a smooth decay, as opposed to the smaller rebound-like oscillations we observe in the UKF and original model. It is unclear which of these dynamics best qualitatively represent the data. We hypothesize that models that best represent underlying dynamics will also best forecast the sparse measurements, but, in practice, simpler models may prove more robust. The parameter estimates from Nelder-Mead effectively simplified the ultradian model in order to minimize the squared error misfit between the sparse data and forecasts.}
  \label{fig:curves}
  \end{adjustwidth}
\end{figure}

In Fig~6, we qualitatively compare the inferred and measured glucose dynamics in P6. Qualitative evaluations of inferred dynamics provide important insight into the physical manifestation of the model fits, and expose how each method balances the observed data with the ultradian glucose model.

We observe that the untrained model (Fig~6a) fails to capture basic statistics of the timeseries, including the mean and variance. The trained models (Fig~6b-6c) better approximate the mean and variance, and are more appropriately calibrated towards equilibirum values.
Oscillations from each of the three fits are noticeably different in amplitude and frequency --- the untrained model exhibits the largest amplitude and oscillation frequency, whereas the UKF dampens these oscillations. The Nelder-Mead parameter estimates fully dampen the nutritional kicks, creating a smooth decay in glucose. 

It is interesting that the Nelder-Mead based forecasts outperform the untrained and dual UKF forecasting in all three metrics for P6, given that the continuous dynamics from Nelder-Mead forecasting lack the oscillations that are clearly present in the implanted glucose sensor data. These oscillations are, if anything, better represented by the untrained and dual UKF approaches. Of course, the performance metrics only compare pointwise forecasts at fingerstick measurement timepoints, which are a sparse sampling from the underlying dynamics.

\subsubsection{Quantitative assessment of inferred continuous dynamics}

\begin{figure}[!h]
    \centering
    \includegraphics[scale=0.3]{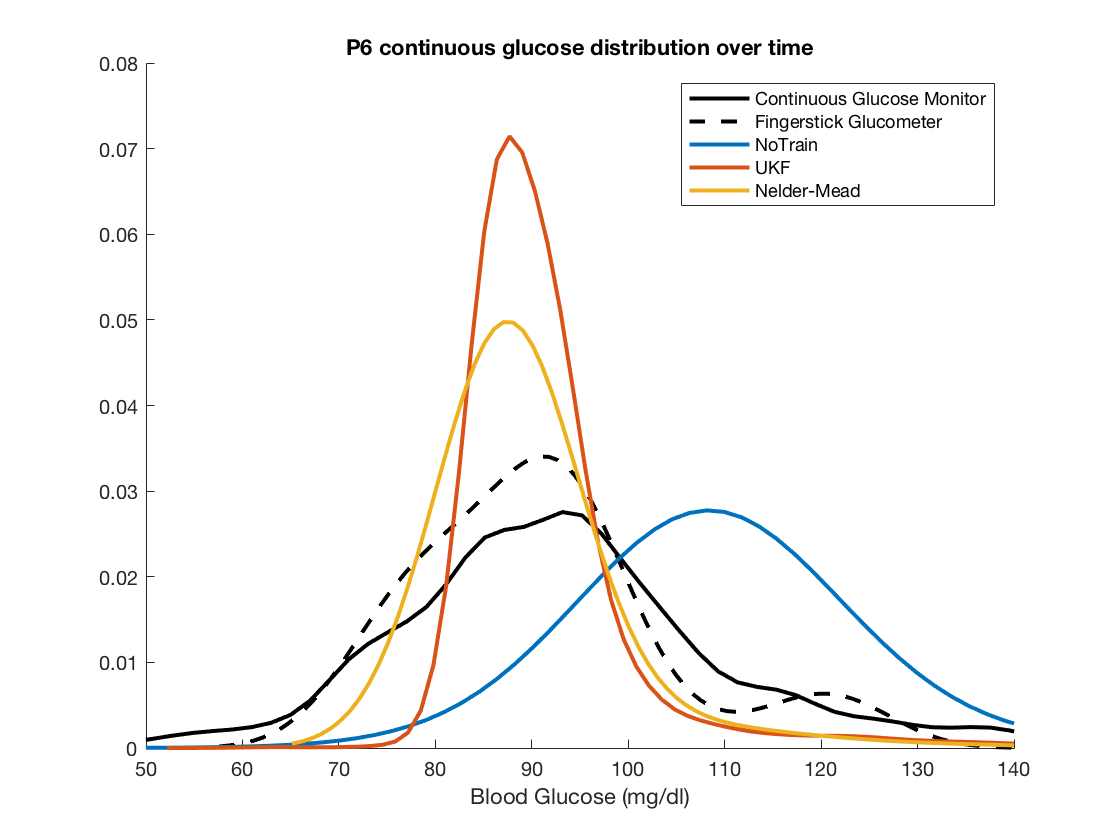}
    \caption{{\bf Reproducing the glucose distribution.} 
    Each inference method and measurement sources capture different statistics of the true underlying glucose distribution. Kernel density estimates of the P6 continuous glucose inferences were computed and plotted for comparison against distributions of P6's continuous and sparse fingerstick measurement distributions. First, we observe that fingerstick and continuous sensor glucoses take on a similar distribution, despite having different sampling frequencies and biases. Fingersticks have a noticeable overrepresentation of low glucose values. The untrained model exhibits poor pointwise predictions in P6, and indeed has a noticeably biased distribution. Interestingly, the KDE of the untrained model dynamics is surprisingly similar to the KDE from continuous measurements, with a large positive shift.}
    \label{fig:KDEcontinuous}
\end{figure}

It is also useful to compare statistics of fingerstick and continuous measurements with statistics of the continuous dynamics created by different approaches to driving, fitting, and filtering the ultradian glucose model. 
Here, we quantitatively compare the distributions of inferred continuous glucose trajectories in P6 for the offline, online, and nominal experiments. Fig~7 shows that the raw fingerstick and continuous monitor distributions are rather similar, suggesting that glucose samples biased towards pre- and post-meal measurements still provide reasonable approximation of the continuous glucose distribution. This gives a window into the types of sampling bias we are likely to see with fingerstick data, which is the current standard-of-care in diabetes self-management. It also suggests that these biases are mild enough that it is possible to recover first and second moments of the glucose distribution from self-monitored fingerstick data. 

Fig~7 also demonstrates that the untrained model is centered around a different mean from the actual data, whereas the UKF and Nelder-Mead based forecasts have more similar means and modes to the raw fingerstick and implanted sensor data. Both the UKF and Nelder Mead approximate the upper tail of the distribution similarly, and generally underestimate the frequency of glucose levels above the mean. Both the UKF and Nelder-Mead forecast dynamics appear to under-represent the lower tail of the sensor and fingerstick glucose distributions--the UKF understimates the low glucose frequency even more than the Nelder-Mead approach. Interestingly, the skew and variance of the untrained model appears quite similar to the true glucose distributions, despite its poor representation of the true mean.



\subsection{Combined online--offline approaches}


\begin{table}[!ht]
\begin{adjustwidth}{-2.25in}{0in} 
\caption{{\bf Best methods per patient.}}
\label{table:bestPerPatient}
\begin{tabular}{|c|l|l|c|l|p{2cm}|p{2cm}|}
\hline
\textbf{Participant} & \textbf{Training Method} & \textbf{Testing Method} & \textbf{MSE} & \textbf{LC} & \textbf{$\%$ Parkes A-zone} & \textbf{Contains Parkes D-E zones} \\ 
\hline \hline
\multicolumn{7}{|l|}{}\\
\multicolumn{7}{|l|}{\textbf{Methods that optimize MSE}}\\
\hline
P1 & MCMC MSE-optimizer & No Filter & 414 & 0.6 & 85 & No \\
P2 & --- & Dual UKF & 1104 & 0.43 & 79.49 & No \\
P3 & Dual UKF & Dual UKF & 234 & 0.71 & 90 & No \\
P4 & MCMC MSE-optimizer & No Filter & 300 & 0.29 & 92.5 & No \\
P5 & MCMC chain(2)-mean & Dual UKF & 413 & 0.1 & 89.74 & No \\
P6 & MCMC chain(4)-mean & No Filter & 190 & 0.33 & 92.5 & No \\
\hline \hline
\multicolumn{7}{|l|}{}\\
\multicolumn{7}{|l|}{\textbf{Methods that optimize Correlation}}\\
\hline 
P1 & Nelder-Mead & State UKF & 7553 & 0.63 & 35.9 & No \\
P2 & --- & Dual UKF & 1104 & 0.43 & 79.49 & No \\
P3 & MCMC chain(5)-mean & State UKF & 1413 & 0.72 & 82.05 & No \\
P4 & MCMC Rcorr-optimizer & Dual UKF & 2590 & 0.43 & 33.33 & No \\
P5 & Dual UKF & Dual UKF & 605 & 0.36 & 62.5 & No \\
P6 & NelderMead & State UKF & 6217 & 0.4 & 51.28 & Yes \\
\hline \hline
\multicolumn{7}{|l|}{}\\
\multicolumn{7}{|l|}{\textbf{Methods that optimize Parkes Quality}}\\
\hline
P1 & MCMC chain(3)-mean & No Filter & 659 & 0.58 & 92.5 & No \\
P2 & MCMC mean1Train & No Filter & 1179 & 0.07 & 87.5 & No \\
P3 & Dual UKF & Dual UKF & 234 & 0.71 & 90 & No \\
P4 & MCMC MSE-optimzer & No Filter & 300 & 0.29 & 92.5 & No \\
P5 & MCMC chain(2)-mean & No Filter & 460 & 0.09 & 92.5 & No \\
P6 & MCMC chain(5)-mean & No Filter & 229 & 0.32 & 95 & No \\
\hline
\end{tabular} \\
\begin{flushleft} The best pairing of offline and online estimation strategies depends on both the patient and the evaluation metric, but more complex Bayesian training methods are typically preferred, and are typically best paired with an unfiltered model. For each evaluation metric, we include a subtable that lists the method that performed best for each data set. We observe that MCMC training with unfiltered forecasting is the most common method to optimize Mean Squared Error. Linear correlation is optimized by a variety of methods for each data-set, and involved dramatic worsenings of MSE and a propensity for clinically significant errors.
\end{flushleft}
\end{adjustwidth}
\end{table}

We wish to to consistently match or outperform previous performance rates, and do this by introducing perturbations of different methodological aspects of training and testing.  We explore new combinations of training and testing methods beyond those tested in Table~\ref{table:simpleOffline}, and introduce fusions of offline parameter estimation with online data assimilation (e.g. learn parameters through Nelder-Mead, and use them in a state UKF). 
For offline parameter estimation, we perform both MCMC and a bounded Nelder-Mead optimization.
For online predictions, we utilize a dual UKF, a state UKF, and an unfiltered forward model with known nutrition. 

Table~\ref{table:bestPerPatient} shows the combination of training and testing methods that resulted in the highest predictive performance for each participant, according to each of our three evaluation metrics.
Results in Table~\ref{table:bestPerPatient} show that MSE is most often optimized by MCMC training. In P2, MSE is minimized by running the dual UKF over the testing set without performing any training --- this is highly indicative of non-stationarity in learned parameters. In P3, MSE is minimized by running the dual UKF over the training and testing sets in sequence. Interestingly, MCMC training is typically best paired with the unfiltered forecasts over the testing set, rather than a prediction-correction  strategy.
The Parkes Quality metric selects the most similar method pairings across individuals, most often choosing MCMC training with deterministic forecasting, except in P3, for whom it selects our established dual UKF approach.
We also observe that linear correlation is a poor stand-alone evaluation metric, since the method pairings that maximized correlation also produce clinically dangerous predictions for four out of the six participants. 

Note that each of the 5 MCMC chains run offline are treated as separate offline training procedures. In some cases, not all MCMC chains exhibit similar performance. For example, the method that optimizes test-set MSE for P5 uses one particular MCMC chain for offline training, and a dual UKF for online forecasting; however, other offline chain means of P5 training data induce much worse performance of the dual UKF, indicating that the chains failed to converge together.

\subsubsection{Average performance across participants}
\begin{table}[!ht]
\begin{adjustwidth}{-2.25in}{0in} 
\caption{{\bf Best methods across participants.}}
\label{table:rankedTable}
\begin{tabular}{|l|l|p{1cm}|p{1cm}|p{1cm}|p{2cm}|p{2cm}|}
  \hline
Training Method & Testing Method & mean $\%A$ Parkes Zone & mean $\%B$ Parkes Zone & mean $\%C$ Parkes Zone & mean $\%$optimal MSE & mean $\%$optimal correlation \\ 
  \hline
  MCMC chain(2)-mean & No Filter & 90.83 & 9.17 & 0.00 & 87.83 & 60.82 \\ 
  MCMC chain(1)-mean & No Filter & 90.00 & 10.00 & 0.00 & 87.32 & 61.83 \\ 
  MCMC MSE & No Filter & 87.92 & 12.08 & 0.00 & 86.68 & 56.15 \\ 
  MCMC chain(4)-mean & No Filter & 87.92 & 12.08 & 0.00 & 85.93 & 58.40 \\ 
  MCMC MAP & No Filter & 87.92 & 12.08 & 0.00 & 84.32 & 57.43 \\ 
  MCMC chain(5)-mean & No Filter & 89.58 & 10.42 & 0.00 & 83.99 & 58.99 \\ 
  MCMC chain(3)-mean & No Filter & 90.42 & 9.58 & 0.00 & 80.28 & 59.15 \\ 
  Nelder-Mead bounded & No Filter & 83.75 & 15.83 & 0.42 & 79.76 & 56.51 \\ 
  Dual UKF & Dual UKF & 76.50 & 23.50 & 0.00 & 64.70 & 54.10 \\ 
  Dual UKF & State UKF & 81.20 & 18.80 & 0.00 & 63.53 & 38.47 \\ 
  --- & Dual UKF & 79.91 & 20.09 & 0.00 & 62.10 & 36.38 \\ 
  Dual UKF & No Filter & 80.83 & 19.17 & 0.00 & 61.80 & 31.85 \\ 
  --- & No Filter & 77.5 & 22.5 & 0.00 & 48.06 & 10.84 \\
  \hline
\end{tabular} \\
\begin{flushleft} Offline training methods paired with unfiltered forecasting in the testing phase outperform strictly online approaches, on average across participants and evaluation criteria.
We average performance measures across data-sets and sort them according to the average degree to which they optimized Mean Squared Error (mean $\%$ optimal MSE). We also report the Parkes error zones and mean $\%$ optimal linear correlation. MCMC-based training methods coupled with unfiltered forecasting performed best overall, followed by Nelder-Mead training with unfiltered forecasting, then dual UKF online filtering. All other combinations of methods (e.g. offline training paired with a filtering online method) create clinically dangerous errors and are excluded from the comparison table. The average improvement from offline methods is best seen concretely by comparing the fraction of errors in the A and B zones of the Parkes Error grid. Offline methods transfer nearly half of the B-zone errors ("altered clinical action, little or no effect on clinical outcome") made by online approaches to the A-zone ("clinically accurate measurements").
\end{flushleft}
\end{adjustwidth}
\end{table}

We wish to identify methods that can work consistently across individuals, since it is beyond the scope of this work to learn the appropriate situations to deploy each method.
To evaluate methods across participants, we average relative error metrics across individuals (Eqs~(\ref{eq:fracMSE}, \ref{eq:fracCorrelation})),
The results of cross-participant method evaluation are presented in Table~\ref{table:rankedTable}, which is ordered by decreasing mean $\%$optimal MSE.

We observe the following key trends: 1) MCMC methods paired with an unfiltered forecasting setup perform best, on average, over all participants, 2) Nelder-Mead methods slightly underperform the MCMC methods, but still outperform the online dual UKF approach in aggregate, 3) the untrained and unfiltered nominal model often leads to clinically dangerous forecasts (not included in Table~\ref{table:rankedTable}), and 4) offline training methods paired with online testing methods often perform even worse than the untrained, unfiltered, nominal model. Nevertheless, the high variance in the applicability of a given method (shown in Table~\ref{table:rankedTable}) across all participants indicates a need for a more generalizable inference scheme or methods for selecting the appropriate inference scheme for each data set.

\subsection{Comparison of parameter estimates}

\begin{figure}[!h]
  \centering
    \begin{adjustwidth}{-2.25in}{0in} 

    \subfloat[MCMC chains]{\includegraphics[scale=0.2]{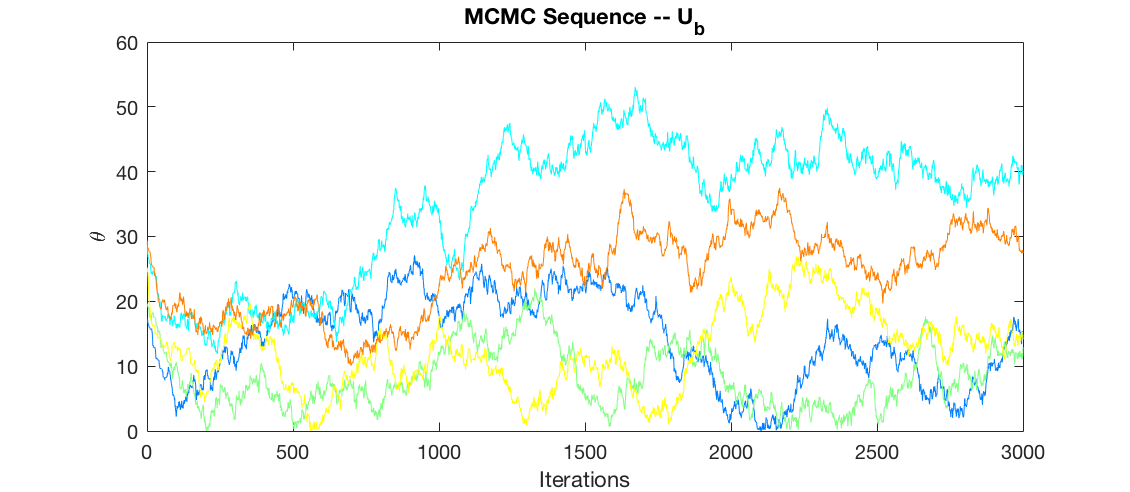}
\label{fig:param_chains}}
\par
  \subfloat[Parameter spread]{\includegraphics[scale=0.45]{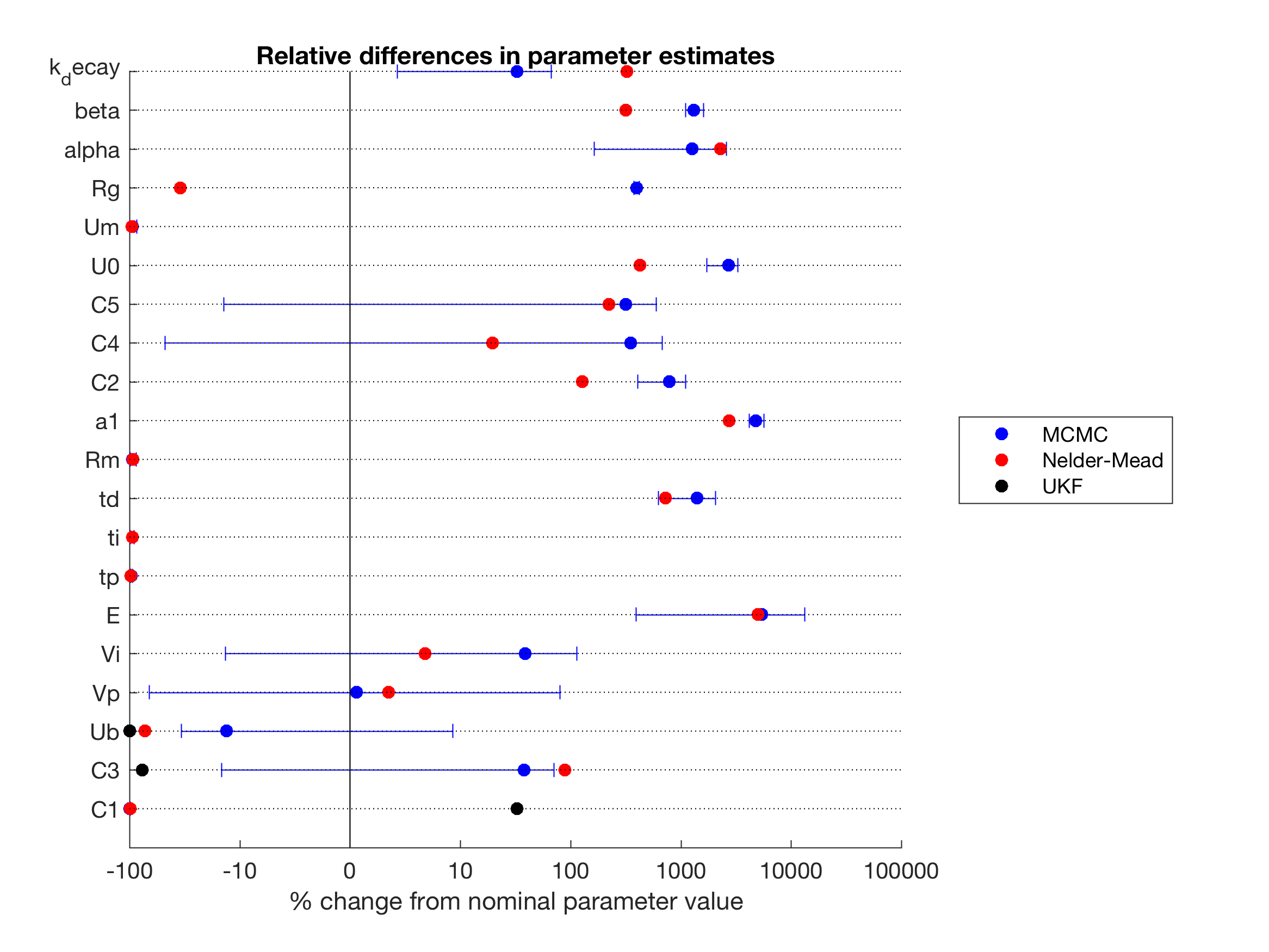}
\label{fig:param_spread}}
  \subfloat[MCMC Densities]{\includegraphics[scale=0.15]{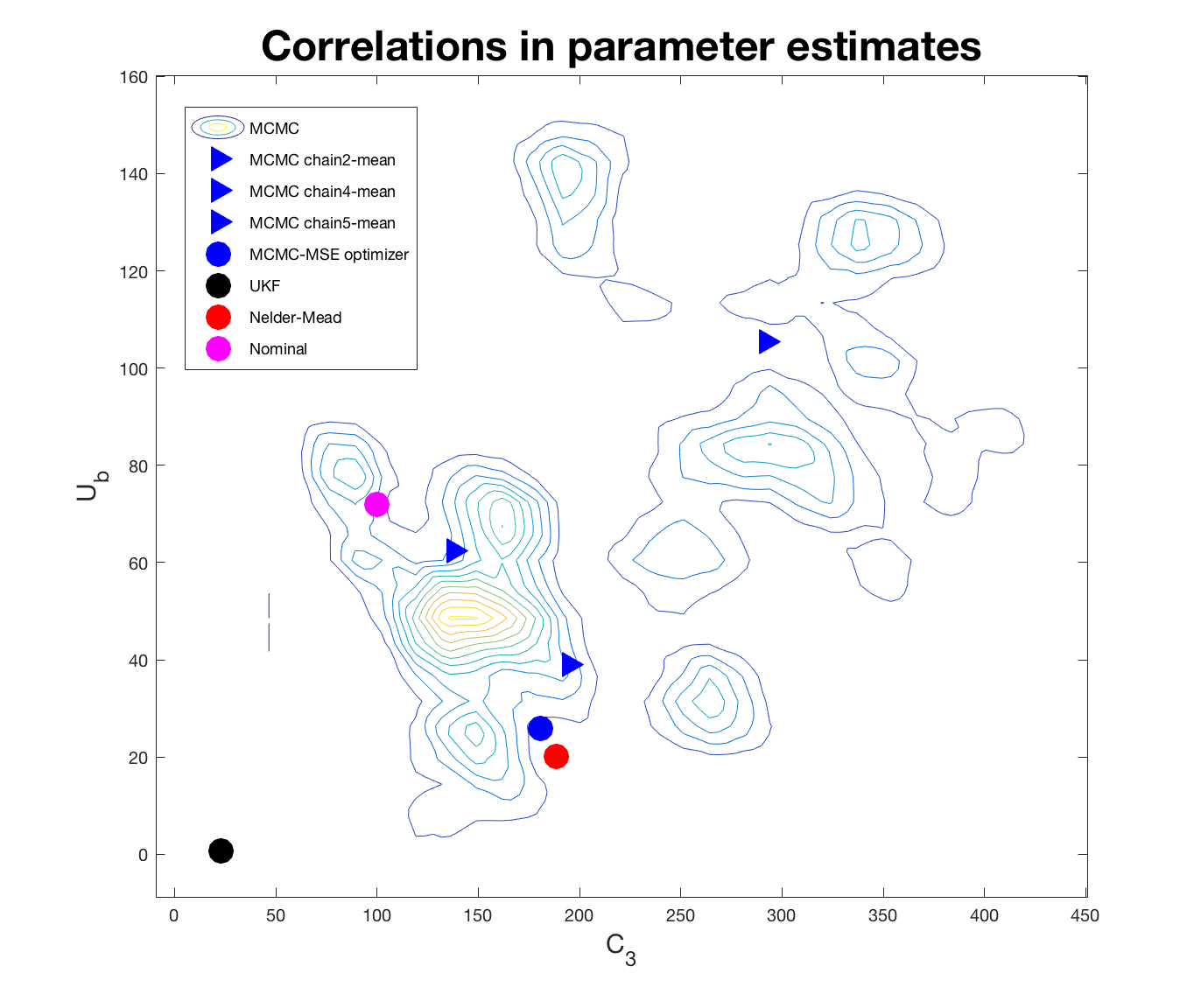}
\label{fig:param_density}}
  \caption{{\bf Parameter estimates from Bayesian Inverse, optimization, and filtering methods.}
  Different parameter estimation methods yield markedly different results, and are each vulnerable to different types of problems. 
  Fig~8a plots the iterations of MCMC for the $U_b$ parameter for 5 independently instantiated chains over the P5 training data. The chains fail to converge to a single consensus distribution, and have high autocorrelation---these are fundamental challenges when working with MCMC.
  Fig~8b shows parameter estimates for P5 generated using MCMC, Nelder-Mead, and a dual UKF in relation to nominal parameter values reported by Sturis et al.. The horizontal error bars represent the $80\%$ quantile of likely parameter values as determined by MCMC, and the red dots indicate point estimates generated through Nelder-Mead optimization. UKF-estimated parameters are represented for the bottom three parameters as black dots. MCMC estimates rather large error bars on some parameters, while finds a tighter fit for others. It is also interesting to note that Nelder-Mead and MCMC sometimes to converge to similar values.
  Fig~8c plots a bivariate density of P5 estimates for $U_b$ and $C_3$, based on the distribution of accepted values across multiple MCMC chains. The means for each chain are overlaied, along with the parameter coordinate that minimized misfit with the training data. We include the UKF and Nelder-Mead estimates, as well as the nominal parameter values for comparison. Clear correlation structure can be observed between $U_b$ and $C_3$. It is also interesting to note that MCMC and Nelder-Mead can function as similar optimizers. Nevertheless, one of the chain means is very far from the rest of the estimates, making clear the challenges with convergence.}
  \label{fig:parameters}
  \end{adjustwidth}
\end{figure}

Here, we directly examine the outputs from the parameter estimation schemes, in order to better understand: 1) the important differences between the parameter estimation schemes, and 2) why online methods perform adequately alone, but degrade when coupled with supposedly improved parameter estimates generated offline. 

Fig~8 presents results from multiple parameter estimation procedures over the P5 training data-set, and indicates the varying degrees to which we are able to achieve convergence and consensus through MCMC, Nelder-Mead, and UKF. Fig~8a demonstrates the challenge of this problem, by showing how 5 independently instantiated chains of MCMC fail to converge upon an estimate for $U_b$, despite many thousands of iterations and 12 hours of computation.

Fig~8b shows the scale of differences between nominal parameter values provided by Sturis et al. and estimates generated when applying different estimation procedures to the P5 training data. For example, we observe that all estimates of $U_b$ are significantly smaller than the nominal value. The estimates of $U_b$ by the dual UKF and Nelder-Mead are rather similar, but still deviate significantly from the expected value (blue circle) and range (blue error bar) determined by MCMC. 
Nelder-Mead and MCMC-based estimates of $C_1$ and $C_3$ were on similar scales, whereas the UKF estimate of these parameters differed wildly from the offline methods. 

Differences in estimates can, in part, be due to correlations across parameters. We can look more closely at the parameter estimates from MCMC in Fig~8c, and see further evidence of large covariance across parameters. Fig~8c also exposes the complexity of the likelihood surface across parameters, which is exacerbated when multiple chains fail to attain the same distribution in the allotted computation time.





\section{Discussion}
We present results that evaluate computational techniques in data assimilation, inverse problems, and physiologic modeling for blood glucose forecasting in the context of real-world self-monitoring data. Recall that progress in this domain can involve addressing challenges in data quality, physiologic modeling, inference methodologies, and evaluation schemes. Here we advance the state of inference methodology by exploring novel combinations of inference methods under real-world data conditions and a single mechanistic model. 

We explore specific contrasting approaches to leveraging mechanistic models for personalized glucose prediction using self-monitoring data collected by people with type 2 diabetes, and find that: 1) an unpersonalized physiologic model can produce dangerous clinical errors in some patients, 2) offline parameter estimation can provide significant improvements over an exclusively online approach, and 3) online (UKF) methods outperform the studied offline methods in an important subset of participants. These results motivate additional study of the relationship between inference schemes and physiologic models in the context of sparse diabetes self-monitoring data, and implicate the need for greater generalizability of inference schemes and methodologies that can select the appropriate inference scheme for each situation.

\subsection{Forecasting without model personalization}
We perform a simple experiment: run the ultradian model simulation forwards with nominal parameters and a fully specified $I_G(t)$ that represents self-monitored carbohydrate intake over the weeks of data; then, compare the simulated glucose trajectory to the self-monitored glucose measurements.
We observe in Table~\ref{table:rankedTable} that an untrained and unfiltered model has substantially greater forecasting errors (and, at times, clinically significant errors) than inference schemes that use data to adjust model parameters and/or track model states. The likelihood of such errors increases with a patient's dissimilarity from the nominal model, and it is unacceptable to allow for such degradation; in fact, these patients are often the ones most in need of targeted therapy and personalization.

It is unsurprising---perhaps obvious, even---to find that an unpersonalized physiologic model is not solely sufficient to produce adequate glucose forecasts. It is well known that glucose-insulin dynamics vary significantly across people, both with and without diabetes~\cite{zeevi_personalized_2015}. Models of the glucose-insulin system are not pre-supposed to represent all people at once, and, instead, often aim to represent average population dynamics. For this reason, the selection of an appropriate inference scheme is paramount. Albers et al.~\cite{albers_personalized_2017} found that, in the context of a UKF, dual tracking of states and parameters of an ultradian model~\cite{sturisUltradian} and of a meal model~\cite{dalla_man} outperformed UKFs that tracked only states or parameters. These findings suggest that model dynamics ought to be personalized to each individual (via parameter estimation), \emph{and} need to be updated (via state estimation) to recover from forecasting errors.

\subsection{Offline parameter estimation with unfiltered forecasting}
The best overall forecasting approach that we test (average across participants of 87\% optimal MSE, 56\% optimal correlation, and 88\% Parkes zone A, see Table~\ref{table:rankedTable}) uses MCMC as an inference scheme for offline parameter estimation over the training set, then performs forecasting over the testing set using a simple unfiltered model (i.e. states and parameters of the ODE system are fixed according to results from MCMC, the time-series of nutrition data are given as $I_G(t)$, and new glucose measurements are ignored). This success demonstrates the value of model-personalization via careful offline parameter tuning.

In principle, MCMC covers the full Bayesian posterior on the unknown states (here parameters in the model) given data, if it converges. 
In fact, we find that when running MCMC on these models over multiple copies in parallel, full statistical convergence is rarely observed. Nonetheless, the algorithm, through working with the model-data misfit, often successfully pushes parameters to regimes where the predictive capability is improved. 
The principle exception to this observation is P2---all methods struggled to fit P2, and parameter proposals in MCMC had very low ($< 0.05$) acceptance probability.
Further tuning of MCMC hyperparameters, including adaptive implementations, would likely improve convergence and thus result in better forecasting across participants.


Given the lack of convergence and the fact that we do not leverage the entire posterior distribution, MCMC is here used primarily as an optimization strategy, rather than a method of uncertainty quantification. Indeed, Nelder-Mead optimization yielded similar, but slightly inferior results to the MCMC-based method we test. Both methods iteratively scan through the parameter space, but only MCMC does this in a way that guarantees eventual convergence of the posterior distribution (with infinite proposal iterations); traditional implementations of Nelder-Mead have no mechanism for escaping locally attractive optima. 
Nevertheless, we find that Nelder-Mead estimates are often quite close to values determined through MCMC (e.g. $C_3$ and $U_b$ in Fig~8c and $E$ in Fig~8b). Still, the two procedures often produce estimates that disagree (e.g. $R_g$ in Fig~8b).

The performance gains from MCMC over Nelder-Mead must be weighed in the context of their computational complexity. Computation time for Nelder-Mead is of order minutes with a single processor, whereas MCMC is run for 12 hours with 5 independently-initialized parallel chains. It is likely that a stochastic optimization routine may offer most of the advantages we get from MCMC in this study.

\subsection{Combining offline parameter estimation and online forecasting}
We hypothesized that an ideal approach to forecasting involves a hybrid of offline parameter estimation and online state tracking. Given a set of initial state and parameter conditions (whether nominal or carefully tuned offline), we expected the unfiltered forecasting model to provide a \emph{lower bound} on the average performance of an unscented Kalman filter. This intuition stems from an assumption that an online filter and an unfiltered model equally benefit from the tuned parameters, and that the online filter creates additional improvements by updating state (and possibly parameter) estimates, given new data.

In the presented experiments, we more often observe the contrary: parameter estimates produced offline, which perform well under unfiltered forecasting (better than untrained, unfiltered models, and typically better than a dual UKF) cause a dramatic decline in performance when they are used to intialize a state UKF. Further investigation is required to better understand this phenomena, but preliminary experiments suggest that offline parameter estimates produce state dynamics that are more sensitive to perturbations. Since a UKF approximates $\hat{x}(t_{k+1})$ by applying the map $\Psi$ to a simplex of perturbations of $\hat{x}(t_k)$, it is unsurprising that increased model sensitivity results in a degradation of forecasting performance.

We find that this model sensitivity is less likely to occur for participants whose data are better represented by the nominal ultradian model. In these cases, online data assimilation is often advantageous for forecasting over the testing set. Combining online filters with offline parameter estimation offered additional advantages only for P5. Interestingly, P5 appears to be best suited to the ultradian model---Table~\ref{table:nominalMSE} in the Appendix shows that the unfiltered nominal ultradian model produces lower MSE for P5 than for other participants, over both the training and testing data. The nominal ultradian model is also well-suited to data from P3, and indeed, forecasts for P3 are optimized when online filtering is performed over the test set.

Together, these results suggest the following: 1) offline estimation procedures are likely able to discover parameter regions that reshape the model to better fit the data, but when these contortions are too great, subsequent data assimilation can be degraded, 2) offline parameter estimates are more compatible with our presently formulated UKF when fewer parameter adjustments are needed (e.g. P5), and 3) UKF is best suited for small parameter and state adjustments in cases where data are already well represented by the model (e.g. P3). 

In order to fully understand the observed degradation of filter performance, we intend to replicate these results in a simpler glucose model (e.g.~\cite{bergman_physiologic_1981}) using simulated data, and will work up to the complexity of enclosed models and data until the phenomena can be duplicated. From there, we will drill down into the specific causes of filter degradation. It is possible that directly addressing model sensitivities in the filter design will prevent these issue and further improve prediction quality. It may be beneficial to consider direct learning of model errors in addition to model parameters in order to preserve the properties of the physiologic models that permit easy use of non-linear stochastic filters. Worthwhile avenues of investigation include additional tuning of hyperparameters and different assumptions about covariance structure and process noises, especially those that leverage higher moments of the posterior distribution produced by MCMC. Other filters may be more suited to dealing with these sensitivities; for example, particle filters do not make gaussian assumptions on the posterior forecast distribution, and may therefore be more robust to misbehavior by a small subset of ensemble members. In addition, there may be opportunities to build constraints or Bayesian priors on the filters' state-space to address these issues~\cite{arnold_parameter_2014}.

\subsection{Model identifiability}
Structural unidentifiabiliity---the existence of infinitely many parameter combinations that produce identical model responses---presents a major challenge for parameter estimation. In addition, structurally identifiable models can be \emph{practically unidentifiable} when parameter differences are imperceptible relative to available data. 
We do not formally evaluate structural identifiability of the ultradian glucose model, but we recognize that its parameters are rarely practically identifiable. For example, combinations of exponential parameters in $f_4(h_3)$ are likely indistinguishable from one another without high frequency, low-noise data. Lack of identifiability can corrupt our belief in parameter estimates, and can also create problematic convergence pathologies in estimation schemes like MCMC.

One of the most common ways to achieve model identifiability is to restrict estimation to an identifiable subset of parameters. We explored this approach, but found that offline-tuned predictions degraded when learning fewer parameters, despite the improvement in identifiability and MCMC convergence. However, we hypothesize that more advanced approaches for dealing with identifiability could provide valuable performance gains. Eisenberg \emph{et al.} have developed and applied novel methods for selecting identifiable combinations of parameters in nonlinear ODEs ~\cite{meshkat_algorithm_2009,eisenberg_identifiability_2013,eisenberg_determining_2014}. These approaches, combined with model reduction techniques, will likely improve predictive performance and, importantly, allow for more trustworthy interpretations of biological parameters and their uncertainties.

\subsection{Considerations for operationalizing a glucose forecasting methodology}
We have deployed a mobile app, GlucOracle, that runs a dual UKF with a mechanistic model to provide users with in-the-moment glucose predictions about a prospectively recorded meal.
Here, we find that a 1-2 week period of intense self-monitoring (majority of meals, with a pre- and post-prandial glucose measurement) can train a mechanistic model better, on average, than the previously validated online approach (dual UKF) by Albers et al.. This may make it possible to relax standards of self-monitoring, and still provide a state-of-the-art forecast, given a current or proposed meal.

Additional considerations for operationalizing an offline approach include defining an appropriate training window (i.e. its size, and whether to define it by number of measurements or clock time) along with the frequency at which parameter estimates are re-calculated (e.g. every week). Additional study is needed to understand how the posterior distribution from one training session would carry forward to inform the prior distribution of the next training session. While MCMC was the best performing training method overall, it is a particularly slow method, and is liable to issues with convergence; thus, practical constraints may favor optimization approaches. Nelder-Mead optimization is orders of magnitude faster and still outperforms the existing dual UKF framework, suggesting that stochastic optimization techniques could be used to more closely approach the performance of MCMC while remaining computationally efficient.

Given the diversity of best-methods observed across 6 participants, it would also be wise to implement model-selection strategies on a per-patient basis. We suspect that a combination of MSE and Parkes quality ratings would provide a reasonable foundation for model selection. Linear correlation is ill-advised, as it often selects methods with excessive MSE and poor Parkes quality. It is also important to understand which methodologies have the fastest learning rate with respect to training set size. We hypothesize that users are more likely to trust and engage with GlucOracle if forecasts are intially reasonable, and gradually improve over time than if they receive poorly calibrated forecasts (or no forecasts) until the training set stabilizes. Ideally, we would perform such a model-selection in real-time, to cope with changing forecast demands and model improvements as more training data become available.

\section{Conclusions}
We advance our understanding of the impact of different inference methodologies on blood glucose forecasting on type 2 diabetes self-monitoring data under a single, ultradian model of glucose dynamics.
We conclude that offline methods for parameter estimation, including Bayesian and optimization approaches, often offer improvements over an existing online data assimilation strategy that uses unscented Kalman filters. In particular, Markov Chain Monte Carlo parameter estimation paired with a simple unfiltered forecasting method was the best method across participants among the methods we tested. 
Unsurprisingly, we do not observe a universally best inference scheme across all participants, and strictly online data assimilation performed best in a subset of participants. More surprisingly, MCMC and other offline training methods perform best when paired not with an online filter, but rather with an unfiltered forecasting model (nutritional drivers are considered, but glucose measurements are ignored). It appears that offline-estimated parameters can degrade filter performance,  providing directions of interesting future study for improving filter stability. Nevertheless, we establish a high-quality forecasting method that may allow for a relaxation of data-capture expectations for individuals engaged in self-monitoring who wish to receive personalized glucose forecasts.


\nolinenumbers


\clearpage

\section{Appendix}

\begin{table}[!ht]
\caption{{\bf MSE of untrained ultradian model over testing and training set}}
\label{table:nominalMSE}
\begin{tabular}{|l|c|c|}
  \hline
Participant & Training & Testing \\ 
  \hline
  P1 & 511 & 796 \\ \hline
  P2 & 2096 & 2331 \\ \hline
  P3 & 589 & 536 \\ \hline
  P4 & 530 & 785 \\ \hline
  P5 & 511 & 541 \\ \hline
  P6 & 595 & 561 \\ \hline
\end{tabular} \\
\begin{flushleft} The ultradian model with nominal parameters is differently suited to the data from each participant. Examining the MSE of the unfit, unfiltered ultradian model over the training and testing set provides insight into how well the nominal ultradian model describes each individual. We see that P3, P5, and P6 are fit best across training and testing data. This also helps to explain why the UKF outperforms offline methods for P3 and P5---it is able to track states properly, because the ultradian model did not require major adjustment to fit the data, which can, in turn, create model instabilities that affect the filter performance. 
\end{flushleft}
\end{table}
\end{document}